\documentclass[12pt]{iopart}
\usepackage{amssymb}
\usepackage{graphicx}
\usepackage{dcolumn}
\usepackage{bm}
\usepackage{epsfig}
\newcommand{\beq}{\begin{equation}}
\newcommand{\eeq}{\end{equation}}
\newcommand{\beqa}{\begin{eqnarray}}
\newcommand{\eeqa}{\end{eqnarray}}
\newcommand{\ba}{\begin{array}}
\newcommand{\ea}{\end{array}}
\usepackage{color}

\begin{document}
\title{Nonlinear quantum model for atomic Josephson junctions
with one and two bosonic species}

\author{Giovanni Mazzarella$^{1}$, Marco Moratti$^{1}$, Luca Salasnich$^{2}$,
 and Flavio Toigo$^{1}$} \address{$^{1}$Dipartimento di Fisica
``Galileo Galilei'' and CNISM, Universit\`a di Padova, Via Marzolo 8, 35131
Padova, Italy
\\$^{2}$CNR-INFM and CNISM, Unit\`a di Padova, Via Marzolo 8, 35131
Padova, Italy
}

\date{\today}

\date{\today}

\begin{abstract}
We study atomic Josephson junctions (AJJs) with one and two
bosonic species confined by a double-well  potential.
Proceeding from the second quantized Hamiltonian, we show
that it is possible to describe the zero-temperature AJJs microscopic
dynamics by means of extended Bose-Hubbard (EBH) models,
which include usually-neglected nonlinear terms.
Within the mean-field approximation, the Heisenberg equations derived from such two-mode models provide
a description of AJJs macroscopic dynamics in terms of ordinary
differential equations (ODEs).  We
discuss the possibility to distinguish
the Rabi, Josephson, and  Fock regimes, in terms of the macroscopic parameters which
appear in the EBH Hamiltonians and, then, in the ODEs.
We compare the predictions for the relative populations of the Bose gases
atoms in the two wells obtained from the numerical solutions of the two-mode
ODEs,  with those deriving from the direct numerical integration of the
Gross-Pitaevskii equations (GPEs). Our investigations
shows that the nonlinear terms of the ODEs are crucial
to achieve a good agreement between ODEs and GPEs approaches,
and in particular to give quantitative predictions of the self-trapping regime.
\end{abstract}

\pacs{03.75.Lm, 03.75.Mn, 03.75.Kk}

\maketitle

\section{Introduction}

The prediction \cite{einstein} of Bose-Einstein condensation (BEC) and
the experimental achievement of BEC \cite{anderson} has played a crucial
role for theoretical and experimental developments in the physics of ultracold atoms.
The study of the atomic counterpart
\cite{leggett,leggett01,smerzi,salasnich,salerno} of the Josephson effect
which occurs in superconductor-oxide-superconductor junctions
\cite{book-barone} - which is an example of macroscopic
quantum coherence - represents one of these developments.
Albiez {\it et al.} \cite{albiez} have provided the first experimental
realization of the atomic Josephson junction (AJJ)
previously analyzed theoretically in a certain number of papers \cite{leggett,leggett01,smerzi,salasnich,salerno}.
In 2007 Gati {\it et al.} \cite{gati}
reviewed the experiment by Albiez { \it et al.}  \cite{albiez}
and compared the experimental data with the
predictions of a many-body two-mode model \cite{milburn} and a mean-field description.
In the above references the analysis of AJJs physics is carried
out in the presence of a single bosonic component.
The possibility to tune intra- and
inter-species interactions \cite{minardi,papp} by means of the Feshbach resonance technique makes possible to study of AJJs with two bosonic species trapped together
by double-well potentials and to use BECs mixtures as powerful instruments to investigate quantum
coherence and nonlinear phenomena, with particular attention to the
existence of self-trapped modes and intrinsically localized states.

In the superfluid regime the dynamics of the relative populations and
relative phases of the Bose condensed atoms can be described by Josepshon's
two-mode equations, which are ordinary differential equations (ODEs), see for example
Refs. \cite{smerzi,xu,satja,diaz1,mazzarella}. This
description is achieved in the presence of a confining double-well potential,
with a single bosonic component \cite{smerzi} and also with bosonic mixtures
\cite{xu,satja,diaz1,mazzarella,diaz2}.  One of the most interesting aspects of AJJs
analysis is to compare the predictions deriving from the
ODEs with the ones obtained from the
Gross-Pitaevskii equations (GPEs). For single component
condensates, Salasnich {\it et al.} \cite{salasnich} have shown
that a good agreement exists between the results obtained from
the GPE and those of the ODEs. Similar agreement was obtained in
\cite{salerno} for AJJs realized with weakly interacting solitons
localized in two adjacent wells of an optical lattice. However,
the situation may be quite different for multicomponent
condensates, due to the interplay of intra- and inter-species
interactions which enlarges the number of achievable states (for
instance, mixed symmetry states can exist only in the presence of the
inter-species interaction) as well as their stability, giving to the
system many more dynamical possibilities. Recently it was shown that for the two components
case the integration of the ODEs allows to predict the analogous of the macroscopic
quantum self-trapping phenomenon observed in AJJs with one bosonic component
\cite{satja,mazzarella}. This
phenomenon has  been discussed
for a two-components nonlinear Schr\"{o}dinger model with a double-well potential
by Wang and co-workers \cite{wang}.  More recently, a comparison between
the reduced ODEs system and the full GPE dynamics was performed, showing that,
for various conditions, a good agreement
exists between the two kinds of predictions \cite{mazzarella}.


The aim of the present work is to analyze how the accuracy of the two-mode
approximation can be improved by taking into account  the usually-neglected
nonlinear terms. These terms derive from the overlaps between wave functions
localized in different  wells. Both for single component and for two components AJJs - introduced in the
second section - we proceed from a full second quantized description  of
the system. In Sec. III we describe the system by
the extended Bose-Hubbard (EBH) Hamiltonian. In the single component case, the EBH Hamiltonian is the two-sites
restriction of the Hamiltonian considered in Refs. \cite{mazz,amico} to
analyze bosons loaded in one dimensional optical lattices. In the two species
case, the EBH Hamiltonian is the extended version of the one considered by Kuklov
and Svistunov in Ref. \cite{svistunov} to study  the counterflow
superfluidity of two-species ultracold atoms.  We note that the study of the two
components bosonic system proceeding from a pure quantum approach is a subject
of wide interest. In fact, this topic is dealt with in certain regions of the
phase space in Ref. \cite{giampaolo} and in the case of
hardcore bosons as discussed in Ref. \cite{roscilde}.

The EBH Hamiltonian sustains the dynamics of the single-particle operators via the Heisenberg equations of motion
\cite{ferrini,averin}.  By performing the mean-field approximation on the
single-particle operators of each component, the improved ODEs are achieved. In
the third section we also discuss how it is possible to distinguish the
Rabi regime, the Josephson regime, and the Fock regime. This analysis is carried out in terms of the macroscopic
parameters involved in the EBH Hamiltonians and, then, at the right hand sides of
the improved ODEs as discussed for single AJJs in Ref. \cite{ananikian}. In Sec.
IV we write down the GPEs for the one and the
two components AJJs. Here we compare the results obtained by numerically integrating the GPEs with the predictions
obtained by numerically solving the improved ODEs. Moreover, in the fourth section we plot the phase-plane portraits of  the dynamical
variables fractional imbalance-relative phase. Finally, in Sec. V we draw our conclusions.


\section{The system}

We consider two interacting dilute and ultracold Bose gases denoted below by $1$ and $2$. We
suppose that the two gases are confined in a double-well trap
produced, for example, by a far off-resonance laser barrier that
separates each trapped condensate in two parts, L (left) and R
(right). We assume, moreover, that the two condensates interact
with each other and that the trapping potential $V_{trap}({\bf
r})$ for both components is taken to be the superposition of a
strong harmonic confinement in the radial ($x$-$y$) plane and of a
double-well (DW) potential in the axial ($z$) direction.
 We model the trapping potential as:
\beq
\label{trap} V_{trap}({\bf r})=\frac{m_i \omega_i^2}{2}
(x^2+y^2)+V_{DW}(z)\; ,\eeq
where $m_i$ is the mass of the $i$th component.
For simplicity we take $\omega_1=\omega_2 \equiv \omega$.
For symmetric configurations in
the $z$ direction, we take - for the $i$th species - the double-well in Eq. (\ref{trap}) as
\begin{eqnarray}
\label{PT} &&V_{DW}(z)=V_{L}(z)+V_{R}(z)\nonumber\\
&&V_{L}(z)=-V_{0}\bigg[Sech^2(\frac{z +z_0}{b})\bigg]\nonumber\\
&&V_{R}(z)=-V_{0}\bigg[Sech^2(\frac{z -z_0}{b})\bigg]\nonumber\\
&&V_{0}=\hbar \omega_i\big[1+Sech^2(\frac{2z_0}{b})\big]^{-1} \;,\nonumber\\
\end{eqnarray} that is the combination of two P\"{o}schl-Teller (PT)
potentials, $V_L(z)$ and $V_R(z)$, centered at the points $-z_0$ and $z_0$,
and separated by a potential
barrier which may be changed by varying $b$ (see Fig. 1).
We use  PT potentials  only for the benefit of
improving accuracy in our numerical GPEs calculations (see the fourth section),
taking advantage of the integrability of the underlying linear
system. We remark, however, that our results apply to a generic double-well potential.
Eigenvalues and eigenfuctions
of the  P\"{o}schl-Teller potential  for a single well  are known analytically.
The wave
functions of the ground state of $V_{\alpha}(z)$ ($\alpha=L,R$),   centered around
$-z_0$ ($+z_0$) are  \cite{landau} :
\begin{eqnarray}
\label{wfpt} &&\phi_{(\alpha,i,PT)}(z)=A\big[1-Tanh^{2}(\frac{z
\pm
z_0}{b})\big]^{B_{i}/2} \nonumber\\
&& B_{i}=-\frac{1}{2}+\sqrt{\frac{2m_{i}V_0
b^2}{\hbar^{2}}+\frac{1}{4}} \; .\end{eqnarray}
The constant  $A$, in Eq. (\ref{wfpt}), ensures the normalization of the wave function in each well.
\begin{figure}[h]
\centerline{
\includegraphics[width=8cm,clip]{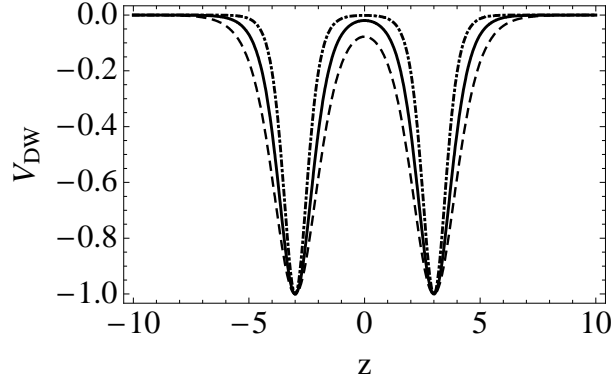}}
\caption{The double-well potential (\ref{PT})  as a function of
$z$ for $z_0=3$ and different values of $b$. The dot-dashed line
corresponds to $b=0.7$, the continuous line corresponds to $b=1$, and the dashed
line corresponds  to $b=1.3$.
Lengths are measured in units of
$\displaystyle{a_{\bot,i}=\sqrt{\frac{\hbar}{m_i\omega}}}$
and energies  in units of
$\hbar \omega$
.}
\label{fig1}
\end{figure}

\section{The second quantization Hamiltonian}
To describe our system at zero-temperature, we proceed from the second quantized Hamiltonian, which
reads
\begin{eqnarray}
\label{sqhamiltonian}
\hat{H} &=&\sum_{i=1,2} \int d^{3} {\bf r}\, \hat{\Psi}^{\dagger}_{i}({\bf r})
\bigg(-\frac{\hbar^2}{2m_i}
\nabla^2+V_{trap}({\bf r})\bigg)\hat{\Psi}_{i}({\bf r})\nonumber\\
&+&\sum_{i=1,2}\frac{g_i}{2}\int d^{3} {\bf r} \hat{\Psi}^{\dagger}_{i}({\bf
r})\hat{\Psi}^{\dagger}_{i}({\bf r}) \hat{\Psi}_{i}({\bf r})
\hat{\Psi}_{i}({\bf
r})\nonumber\\
&+&g_{12}\int d^{3} {\bf r} \hat{\Psi}^{\dagger}_{1}({\bf
r})\hat{\Psi}^{\dagger}_{2}({\bf r})\hat{\Psi}_{2}({\bf
r})\hat{\Psi}_{1}({\bf
r})\;, \end{eqnarray}
where $V_{trap}({\bf r})$ is the potential (\ref{trap}).  The coupling
constants $g_i$ and $g_{12}$ are the intra- and inter-species atom-atom
interaction strengths, respectively. These constants are given by
\begin{eqnarray}
\label{gicoupling} g_{i}=\frac{4 \pi \hbar^{2} a_i}{m_i}\;,
\end{eqnarray}
\begin{eqnarray}
\label{gijcoupling} g_{12}=\frac{2 \pi \hbar^{2} a_{12}}{m_r}
\;,\end{eqnarray} where the reduced mass $m_r$ is equal to
$m_{1}m_{2}/(m_1+m_2)$. Eqs. (\ref{gicoupling}) and (\ref{gijcoupling})
relate the two coupling constants to the respective s-wave
scattering lengths, $a_{i}$ and $a_{12}$. In the following, we shall consider
both $g_i$ and $g_ {12}$ as free parameters, due to the possibility of
changing the s-wave scattering lengths $a_i$ and $a_{12}$ by the
technique of Feshbach resonances.
In the following, we will neglect the mass difference between the two bosonic
components of the mixture, as for example in Ref. \cite{papp}, and assume that $m_1=m_2 \equiv m$.
In Eq. (\ref{sqhamiltonian}), the field $\hat{\Psi}_{i}({\bf r})$ ($\hat{\Psi}^{\dagger}_{i}({\bf r})$)
destroys (creates) a boson of the $i$th species at the point ${\bf
r}$, and obeys  the usual bosonic commutation relations.
We expand
the field operator $\hat{\Psi}_{i}({\bf r})$ in terms of operators
$\hat{a}_{\alpha,i}$ ($\hat{a}^{\dagger}_{\alpha,i})$ - destroying (creating)
a boson of the $i$th species
in the well $\alpha=L,R$ - according to:
\begin{eqnarray}
\label{expansion}
&&\hat{\Psi}_{i}({\bf r})=\sum_{\alpha=L,R}
\Phi_{\alpha,i}({\bf r})\hat{a}_{\alpha,i}\;,\end{eqnarray}
where $\hat{a}$'s  and $\hat{a}^{\dagger}$'s satisfy the usual boson commutation
relations
and the functions $\Phi_{\alpha,i}$  form an orthonormal set.
Due to the form (\ref{trap}) of the trapping potential, $\Phi_{\alpha,i}({\bf
r})$ can be decomposed as
\begin{eqnarray}
\label{decomposition}
\Phi_{\alpha,i}({\bf
r})=w_i(x)w_i(y)\phi_{\alpha,i}(z)\; , \end{eqnarray}
where $w_i(x)$ and $w_i(y)$ are the ground state wave
functions of the harmonic oscillator potentials
$\displaystyle{m_i\omega_i^2x^2/2}$ and
$\displaystyle{m_i\omega_i^2y^2/2}$, respectively.
The functions $\phi_{L,i}(z)$ and $\phi_{R,i}(z)$ at right hand side of Eq.
(\ref{decomposition}) are two functions well localized in the left and right well, respectively.  These functions are
real and orthonormal.
The functions $\phi_{L,i}(z)$ and $\phi_{R,i}(z)$ can be determined following the same perturbative
approach as in Ref. \cite{mazzarella}.
Under the same conditions, these functions may be written in terms of the $\phi_{(L,i,PT)}(z)$ and $\phi_{(R,i,PT)}(z)$ of Eq. (\ref{wfpt}) as:
\begin{eqnarray}
\label{finaldecomposition} &
\phi_{L,i}(z)=\frac{1}{2}\bigg[(\frac{1}{\sqrt{1+s}}+\frac{1}{\sqrt{1-s}})\phi_{(L,i,PT)}(z)+(\frac{1}{\sqrt{1+s}}-\frac{1}{\sqrt{1-s}})\phi_{(R,i,PT)}(z)\bigg] \nonumber\\
&
\phi_{R,i}(z)=\frac{1}{2}\bigg[(\frac{1}{\sqrt{1+s}}-\frac{1}{\sqrt{1-s}})\phi_{(L,i,PT)}(z)+(\frac{1}{\sqrt{1+s}}+\frac{1}{\sqrt{1-s}})\phi_{(R,i,PT)}(z)\bigg]
\
\;, \nonumber\\
\end{eqnarray}
where $s=\int_{-\infty}^{+\infty}dz\, \phi_{(L,i,PT)}(z)
\phi_{(R,i,PT)}(z)$.

\subsection{AJJs with a single bosonic species}

Let us start our analysis by considering the presence of  a single
bosonic component. In this case, the inter-species coupling
constant (\ref{gijcoupling}) is equal to zero.  We use the field operator
expansion (\ref{decomposition}) in the second quantized Hamiltonian
(\ref{sqhamiltonian}). The AJJs microscopic dynamics
is controlled by the EBH Hamiltonian \cite{mazz,ferrini,averin}. The
EBH model, by omitting the species index $i$, is described by the Hamiltonian
\begin{eqnarray}
\label{bhsingle} \hat{H}_{EBH} &=& E_L^{0}
\hat{a}_L^{\dagger}\hat{a}_L+E_R^{0}
\hat{a}_R^{\dagger}\hat{a}_R+\frac{U_L}{2}\hat{a}_L^{\dagger}\hat{a}_L^{\dagger}\hat{a}_L\hat{a}_L\nonumber\\
&+& \frac{U_R}{2}\hat{a}_R^{\dagger}\hat{a}_R^{\dagger}\hat{a}_R\hat{a}_R
-K\big(\hat{a}_L^{\dagger}\hat{a}_R+\hat{a}_R^{\dagger}\hat{a}_L\big)\nonumber\\
&+&K_c\big(\hat{a}_L^{\dagger}\hat{n}_L\hat{a}_R+\hat{a}_L^{\dagger}\hat{n}_R\hat{a}_R+\hat{a}^{\dagger}_R\hat{n}_L\hat{a}_L+
\hat{a}_R^{\dagger}\hat{n}_R\hat{a}_L\big)\nonumber\\
&+& V\hat{a}_L^{\dagger}\hat{a}_R^{\dagger}\hat{a}_L\hat{a}_R+K_p\big(
\hat{a}_L^{\dagger}\hat{a}_L^{\dagger}\hat{a}_R\hat{a}_R
+\hat{a}_R^{\dagger}\hat{a}_R^{\dagger}\hat{a}_L\hat{a}_L\big) \;.  \nonumber\\
\end{eqnarray}
Here $\hat{n}_{\alpha}=\hat{a}^{\dagger}_{\alpha}\hat{a}_{\alpha}$ is the number of
particles in the $\alpha$th well. $E_{\alpha}^{0}$ are the energies of the two wells, $U_{\alpha}>0$
are the boson-boson repulsive interaction amplitudes, and $K$ is
the tunnel matrix element, which is the Rabi oscillation energy in the case of a
model with $U_{\alpha}$ equal to zero. The parameter $K_c$ is the induced
collisionally hopping amplitude, $V$ is the density-density bosonic
interaction amplitude, and $K_p$ describes the pair bosonic hopping \cite{mazz}.
By using the decomposition (\ref{decomposition}) and the explicit form of
$w(x)$ and $w(y)$, the macroscopic parameters (\ref{bhsingle}) may be shown to be related to the intra-species coupling
constant (\ref{gicoupling}) and to the other microscopic parameters (the mass
and the frequency of the harmonic trap) by the formulas
\begin{eqnarray}
\label{selfparameters} && E_{\alpha}^{0}=\int dz \bigg[
\frac{\hbar^2}{2m}(\frac{d\phi_{\alpha}}{dz})^{2}+(V_{DW}
+\frac{\hbar^2}{2m a^{2}_{\bot}}+\frac{m \omega^2
a^{2}_{\bot}}{2})(\phi^{\alpha})^2\bigg]\nonumber\\
&&U_{\alpha} =\tilde g\int_{-\infty}^{+\infty}{dz\,({\phi_{\alpha}}(z))^4}\nonumber\\
&&K=-\int dz \,\bigg[\frac{\hbar^2}{2m}\frac{d \phi_{L}}{d
z}\frac{d \phi_{R}}{d z}
+ V_{DW}(z)\phi_{L}\phi_{R}\bigg]\nonumber\\
&&K_c= \tilde
g\int_{-\infty}^{+\infty}{dz\,({\phi_{\alpha}}(z))^3\,\phi_{\beta}(z)}\nonumber\\
&&V= 2\tilde
g\int_{-\infty}^{+\infty}{dz\,({\phi_{\alpha}}(z))^2\,({\phi_{\beta}}(z))^2}\nonumber\\
&&K_p=\frac{V}{4} \;,
\end{eqnarray}
where $\tilde g=\displaystyle{\frac{g}{2\pi a^2_{\bot}}}$. We observe that the
first two lines of Hamiltonian (\ref{bhsingle})  involve
only the overlaps between $\phi_{\alpha}$'s localized in the same well, see \cite{smerzi}. The third and fourth lines
of the Hamiltonian (\ref{bhsingle}) include also the overlaps between $\phi_{\alpha}$'s localized in different wells, see
\cite{ananikian}. Proceeding from the Hamiltonian (\ref{bhsingle}), we write down criteria to individuate different oscillations
regimes sustained by the AJJs dynamics. To this end, as discussed in Refs.
\cite{leggett01,ananikian}, we express the Hamiltonian (\ref{bhsingle}) in terms of the
following operators:
\begin{eqnarray}
\label{spinsingle}
&&\hat{J}_{x}=\frac{1}{2}(\hat{a}^{\dagger}_{L}\hat{a}_{L}-\hat{a}^{\dagger}_{R}\hat{a}_{R})\nonumber\\
&&\hat{J}_{y}=\frac{i}{2}(\hat{a}^{\dagger}_{L}\hat{a}_{R}-\hat{a}^{\dagger}_{R}\hat{a}_{L})\nonumber\\
&&\hat{J}_{z}=\frac{1}{2}(\hat{a}^{\dagger}_{L}\hat{a}_{R}+\hat{a}^{\dagger}_{R}\hat{a}_{L})
\;,\end{eqnarray}
and the $SU(2)$ algebra invariant $\hat{J}^2=(\hat{N}/2)(\hat{N}/2+1)$,
with $\hat{N}$ being equal to $\hat{n}_L+\hat{n}_R$ \cite{definition}.

We assume that the two potential wells are
symmetric, $E^{0}_L=E^{0}_R \equiv E$ and $U_L=U_R \equiv U$. Neglecting constant terms, and using the fact that $N \gg 1$, we
get
\beq
\label{spinhamiltoniansingle}
\hat{H}=(U-V)\hat{J}^{2}_{x}-2\,(K-K_c\,N)\hat{J}_{z}+V\hat{J}^{2}_{z}
\;.\eeq
Here, if the condition $(U-V) \gg V$ is verified, we can consider only the terms in
$\hat{J}^{2}_{x}$ and $\hat{J}_{z}$. Then by defining the parameter $R$
as
\beq
\label{erre}
R=\frac{(U-V)\,N}{(K-K_c N)}
\;\eeq
we are able (see Ref.
\cite{leggett01}) to distinguish the three following regimes
\begin{itemize}
\item
Rabi: $R \ll 1$;
\item
Josephson: $1\ll R \ll N^2$;
\item
Fock: $N^2 \ll R $.
\end{itemize}
In the Rabi regime the bosons are in a coherent state
and oscillate with a frequency given simply by the energy difference between the ground state
and the first excited state associated to the double-well potential. In the Josephson regime the bosons are in a coherent state
and oscillate with a frequency which depends on the parameters $U$, $K_c$, and
$V$. Moreover, if the interaction strength is sufficiently
large the self-trapping takes place. In the Fock regime the bosons are in a Fock state
characterized by the suppression of number fluctuations.
Now, we observe that the Hamiltonian (\ref{bhsingle}) can be viewed as the two sites restriction of
the Hamiltonian considered in Refs. \cite{mazz,amico} within the study of
bosons loaded in one dimensional optical lattices. In particular, in Ref. \cite{mazz} it is shown that within the Fock regime two
regions open up. To this end we denote by $\Delta$
the energetic gap between the Fock state with $N_0$ bosons per well and the Fock
state with $(N_0+1)$ per well \cite{mazz}
\beq
\label{delta}
\Delta=2(E+U\,N_0)
\;.\eeq
Then, when
\beq
|4\Delta-(2\,N_{0}+1)\,V|>V
\eeq
we have a pure Mott insulating phase (PMI), driven by the density-density
on-site interaction. When
\beq
|4\Delta-(2\,N_{0}+1)\,V|<V
\eeq
we have a Density-Wave Mott insulating (DWMI) regime, driven by the
nearest-neighbors interaction \cite{mazz}. Note that the DWMI phase is characterized by number fluctuations suppression as well.



At this point, we remark that we are interested in determining the fully
coherent dynamical oscillations of
population of the Bose condensed atoms between the left and right
wells.
Then, we proceed from the Heisenberg equations of motion for the model
Hamiltonian (\ref{bhsingle}). These equations of motion control the temporal evolution of $\hat{a}_{\alpha}$.
We observe that in the superfluid regime the system is in
a coherent state and the following mean-field approximation \cite{ferrini}
\begin{eqnarray}
\label{mf}
&&\langle \hat{a}_{\alpha}\rangle=\sqrt{N_{\alpha}}\exp(i\theta_{\alpha}) \nonumber\\
&& \langle \hat{n}_{\alpha}\rangle=N_{\alpha} \;\end{eqnarray}
can be performed. The averages involved in Eq. (\ref{mf}) are evaluated with
respect to the coherent state. Under the assumption of symmetric wells and by
inserting the mean-field approximation (\ref{mf}) into the aforementioned Heisenberg equations of motion, we get
\begin{eqnarray}
\label{singlecomplete}
&&\dot{z}(t)=
-\frac{2(K-K_cN)}{\hbar}\,\sqrt{1-z^2(t)}\,\sin\theta(t)\nonumber\\
&+&\frac{VN}{2\hbar}(1-z^2(t))\sin2\theta(t)\nonumber\\
&&\dot{\theta}(t)=
\frac{U-V}{\hbar}N z(t)+\frac{2(K-K_cN)}{\hbar}\,\frac{z(t)\cos\theta(t)}{\sqrt{1-z^2(t)}}\nonumber\\
&-&\frac{VN}{2\hbar}\,z(t)\cos2\theta(t)\;,
\end{eqnarray}
where $N=N_L+N_R$ is the total number of
bosons, and $z=(N_L-N_R)/N$ and $\theta=\theta_R-\theta_L$  are, respectively, the fractional
imbalance and the relative phase.

\subsection{AJJs with two bosonic species}
In this subsection we shall consider AJJs in the presence of two interacting bosonic components.
In this case both the coupling constants (\ref{gicoupling}) and (\ref{gijcoupling}) are
finite, and the two mode EBH model is described by the
Hamiltonian
\beq
\label{bhtwo}
\hat{H}=\sum_{i=1,2}\hat{H}_{(EBH,i)}+\hat{H}_{12}\; .
\eeq
The Hamiltonian $\hat{H}_{(EBH,i)}$ is the single component Hamiltonian
(\ref{bhsingle}) written in terms of the operators $\hat{a}_{\alpha,i}$ and
$\hat{a}^{\dagger}_{\alpha,i}$. The
parameters $E^{0}_{\alpha}$, $U_{\alpha}$, $K$, $K_c$, $V$, $K_p$, and the function $\phi_{\alpha}$ will read
$E^{0}_{\alpha,i}$,  $U_{\alpha,i}$, $K_i$, $K_{c,i}$, $V_i$, $K_{p,i}$, and
$\phi_{\alpha,i}$, respectively. The microscopic quantities referred to a single
bosonic component will be modified according to the same prescription.
Under the hypothesis of symmetric wells, the coupling
Hamiltonian $\hat{H}_{12}$ reads:
 \begin{eqnarray}
 \label{12hamiltonian}
 \hat{H}_{12} &=&
U_{12}\big(\hat{a}^{\dagger}_{L,1} \hat{a}^{\dagger}_{L,2}
 \hat{a}_{L,1} \hat{a}_{L,2}+\hat{a}^{\dagger}_{R,1} \hat{a}^{\dagger}_{R,2}
 \hat{a}_{R,1} \hat{a}_{R,2}\big)\nonumber\\
&+& V_{12}\big(\hat{a}^{\dagger}_{L,1} \hat{a}^{\dagger}_{R,2}
 \hat{a}_{L,1} \hat{a}_{R,2}+\hat{a}^{\dagger}_{L,2} \hat{a}^{\dagger}_{R,1}
 \hat{a}_{L,2} \hat{a}_{R,1} \big)\nonumber\\
&+& K_{p,12} \big(\hat{a}^{\dagger}_{L,1} \hat{a}^{\dagger}_{L,2} \hat{a}_{R,2}
\hat{a}_{R,1}+\hat{a}^{\dagger}_{R,1} \hat{a}^{\dagger}_{R,2} \hat{a}_{L,2}
\hat{a}_{L,1}\nonumber\\
&+&\hat{a}^{\dagger}_{L,1} \hat{a}^{\dagger}_{R,2} \hat{a}_{L,2}
\hat{a}_{R,1}+\hat{a}^{\dagger}_{R,1} \hat{a}^{\dagger}_{L,2} \hat{a}_{R,2}
\hat{a}_{L,1}\big)\nonumber\\
&+&K_{c,12}\big(\hat{a}^{\dagger}_{L,1}\hat{n}_{L,2}\hat{a}_{R,1}+\hat{a}^{\dagger}_{R,1}\hat{n}_{L,2}\hat{a}_{L,1}
\nonumber\\
&+&\hat{a}^{\dagger}_{L,2}\hat{n}_{L,1}\hat{a}_{R,2}+\hat{a}^{\dagger}_{R,2}\hat{n}_{L,1}\hat{a}_{L,2}\nonumber\\
&+&
\hat{a}^{\dagger}_{L,2}\hat{n}_{R,1}\hat{a}_{R,2}+\hat{a}^{\dagger}_{R,2}\hat{n}_{R,1}\hat{a}_{L,2}\nonumber\\
&+&
\hat{a}^{\dagger}_{L,1}\hat{n}_{R,2}\hat{a}_{R,1}+\hat{a}^{\dagger}_{R,1}\hat{n}_{R,2}\hat{a}_{L,1}\big)
\; .\end{eqnarray}

In Eq. (\ref{12hamiltonian}), $U_{12}$ is the inter-species interaction
amplitude between bosons localized in the same well, and $V_{12}$ is the inter-species interaction
amplitude between bosons localized in different wells. The quantity $K_{p,12}$
is the inter-species pair hopping (hopping of particle-particle or hole-hole pair
made up of bosons of different species); $K_{c,12}$ is the amplitude of the inter-species collisionally induced hopping.
By using the decomposition (\ref{decomposition}) and the explicit form of
$w_i(x)$ and $w_i(y)$,
the aforementioned parameters are shown to be related to the inter-species
coupling constant (\ref{gijcoupling}) by:
\begin{eqnarray}
\label{12selfparameters}
&&U_{12}=\tilde{g}_{12}\int_{-\infty}^{+\infty}{dz\,({\phi_{\alpha,i}(z)})^2({\phi_{\alpha,j}(z)})^2}\nonumber\\
&&V_{12}=\tilde{g}_{12}\int_{-\infty}^{+\infty}{dz\,({\phi_{\alpha,i}(z)})^2(\phi_{\beta,j}(z)})^2\nonumber\\
&&K_{c,12}=\tilde{g}_{12}\int_{-\infty}^{+\infty}\,dz\,
(\phi_{\alpha,i}(z))^{3}(\phi_{\beta,j}(z))
\nonumber\\
&&K_{p,12}=V_{12}\; ,
\end{eqnarray}
where $\displaystyle{\tilde g_{12}=\frac{g_{12}}{\pi
(a^{2}_{\bot,1}+a^{2}_{\bot,2})}}$.  Note that we are considering both the overlaps
between $\phi_{\alpha}$'s localized in the same well
($U_{\alpha,i}$ and $U_{12}$) - that are the only terms taken
into account in Ref. \cite{mazzarella} - and the overlaps between $\phi_{\alpha}$'s localized in
different wells ($V_{12}$, $K_{p,12}$, $K_{c,12}$). We observe that, in general, due to the presence of the parameters
(\ref{12selfparameters}) the identification of different
oscillation regimes proceeding from the Hamiltonian (\ref{bhtwo}) is not immediate as for single component AJJs. Nevertheless,
under certain conditions we are able to write down criteria to select the different
regimes sustained by the two components AJJs dynamics.
First, let us focus on the case in which only the overlaps between
$\phi_{\alpha}$'s localized in the same well are considered. If certain relations exist between the
intra- and the inter-species interactions amplitudes, we
can recognize the two-species corresponding of the Rabi, Josephson and Fock
regimes discussed in the case of single component AJJs. For each component $i$, we define the quantity $\gamma_i$ as
\beq
\label{gamma}
\gamma_i=\frac{U_i N_i}{K_i}
\; . \eeq
We recognize the following "weak-coupled" Rabi, Josephson, and Fock regimes
\begin{itemize}
\item
Rabi: $ \gamma_i \ll 1$, $|U_{12}| \simeq U_i$;
\item
Josephson: $1\ll \gamma_i \ll N_{i}^2$, $|U_{12}| \le U_i$;
\item
Fock: $N_{i}^2 \ll \gamma_i $.
\end{itemize}
In the Josephson regime, even if the intra-species interaction is not  strong enough to ensure self-trapping by itself, self-trapping occurs
 when the inter-species interaction strength exceeds a crossover value.
In the Fock regime the net number of atoms in the transport is suppressed.
However, with repulsive inter-species interaction the so-called counterflow
survives \cite{svistunov}. This means that the currents of the two species are equal in absolute
values and are in opposite directions. This conductive regime is named
super(counter)fluid phase (SCF). As discussed in Ref. \cite{svistunov}, the system
supports the SCF phase of the two components when
\beq
\label{scf}
U_1+U_2- 2\,U_{12} \gg1
\;.\eeq
When the condition
\beq
\label{ps}
U_1+U_2=2\,U_{12}
\eeq
is met, a phase separation (PS) is observed in the system and
the system can be viewed as composed by two totally independent Bose gases confined in the
double-well potential. On a physical level, this phase separation means
that one bosonic component will occupy the left well and the other the right well.
If the inter-species interaction is attractive and the hypothesis $N_1=N_2 \equiv
N$ is verified, then, when
\beq
\label{spf}
U_1+U_2-2\,|U_{12}| \gg 1
\eeq
a superfluid phase, in which the superfluid consists of pairs
of bosons, is supported by the system. This phase is named superfluid paired phase \cite{svistunovp}.

So far we have neglected the role played by the terms deriving from the
overlaps between $\phi_{\alpha}$'s localized in different wells. The presence of
these terms makes the scenario more complicated. However, also in this situation, under certain
conditions, it is possible to achieve a classification of the oscillations
regimes.  To this end, as discussed for the single component case, we express the Hamiltonian (\ref{bhtwo}) in terms of the
operators $\hat{J}_{x,i}$, $\hat{J}_{y,i}$, $\hat{J}_{z,i}$ defined in Eq. (\ref{spinsingle})
and the $SU(2)$ algebra invariant $\hat{J}_{i}^2=(\hat{N}_{i}/2)(\hat{N}_{i}/2+1)$,
with $\hat{N}_{i}$ being equal to $\hat{n}_{L,i}+\hat{n}_{R,i}$. Since we are assuming
symmetric potential wells, we can write that $E^{0}_{L,i}=E^{0}_{R,i} \equiv E_{i}$, $U_{L,i}=U_{R,i} \equiv
U_i$. Neglecting constant terms, and using the fact that $N_i \gg 1$, we
get
\begin{eqnarray}
\label{spinhamiltoniantwo}
&&\hat{H}=
(U_i-V_i)\hat{J}^{2}_{x,i}-2\,(K-K_{c,i}\,N_i-K_{c,12}\,N_j)\hat{J}_{z,i}\nonumber\\
&+&V_{i}\hat{J}^{2}_{z,i}+4\big((U_{12}-V_{12})\hat{J}_{x,1}\hat{J}_{x,2}+V_{12}\hat{J}_{z,1}\hat{J}_{z,2}\big)
\nonumber\\
&+&U_{12}(\hat{n}_{L,1}\hat{n}_{R,2}+\hat{n}_{R,1}\hat{n}_{L,2})+V_{12}(\hat{n}_{L,1}\hat{n}_{L,2}+\hat{n}_{R,1}\hat{n}_{R,2})\;.
\nonumber\\ \end{eqnarray}
Again, if $(U_i-V_i) \gg V_{i}, V_{12}$, and $(U_{12}-V_{12}) \gg V_{i}, V_{12}$, we
can consider  only the terms in $\hat{J}^{2}_{x,i}$, $\hat{J}_{z,i}$, and
$\hat{J}_{x,1}\hat{J}_{x,2}$. We will assume also that $N_1=N_2
\equiv N$, $U_1=U_2 \equiv U$, $V_1=V_2
\equiv V$, and $K_{c,1}=K_{c,2} \equiv K_{c}$, and that the initial conditions are the same for both the components.
In analogy to the case of a single component AJJ, we define the parameter $\tilde R$ as
\beq
\label{erretilde}
\tilde R=\frac{\big((U-V)+4(U_{12}-V_{12})\big)\,N}{\big(K-(K_c+K_{c,12})N\big)}
\;.\eeq
Again, we are able to distinguish the three regimes:
\begin{itemize}
\item Rabi: $ \tilde R \ll 1$;
\item Josephson: $1\ll \tilde R \ll N^2$;
\item Fock: $N^2 \ll \tilde R $.
\end{itemize}

At this point, we remark that we are interested in determining the fully
coherent dynamical oscillations of population of the two bosonic components between the left and right
wells. Then, we proceed from the Heinseberg equations of motion for the
model Hamiltonian (\ref{bhtwo}). These equations of motion control the temporal evolution of $\hat{a}_{\alpha,i}$.
Again, by inserting the mean-field approximation valid in the superfluid regime
- $\langle \hat{a}_{\alpha,i}\rangle=\sqrt{N_{\alpha,i}}\exp(i\theta_{\alpha,i}) $,
$ \langle \hat{n}_{\alpha,i}\rangle=N_{\alpha,i} $ -
into the aforementioned Heisenberg equations
of motion, one gets the coupled differential equations for the fractional imbalance
 $z_i=(N_{L,i}-N_{R,i})/N_i$  and  relative phase $\theta_i=\theta_{R,i}-\theta_{L,i}$ of the two species:
\begin{eqnarray}
\label{twocomplete}
\dot{z}_i(t)
&=&-\frac{2(K_i-K_{c,i}N_i)}{\hbar}\,\sqrt{1-z_i^2(t)}\,\sin\theta_i(t)\nonumber\\
&+&\frac{
V_{i}N_i}{2\hbar}(1-z_i^2(t))\sin2\theta_i(t)\nonumber\\
&+&\frac{2}{\hbar}(V_{12}\sqrt{1-z_j^2(t)}\,\cos\theta_j(t) \nonumber\\
&+&K_{c,12})N_j\sqrt{1-z_i^2(t)}\,\sin\theta_i(t)
\nonumber\\
\dot{\theta}_i(t) &=&
\frac{U_i-V_i}{\hbar}N_iz_i(t)+\frac{2(K_i-K_{c,i}N_i)}{\hbar}\,
\frac{z_i(t)\cos\theta_i(t)}{\sqrt{1-z_i^2(t)}}
\nonumber\\
&-&\frac{V_{i}N_i}{2\hbar}\,z_i(t)\cos2\theta_i(t)+\frac{U_{12}-V_{12}}{\hbar}N_jz_j(t)\nonumber\\
&-&\frac{2}{\hbar}(V_{12}\sqrt{1-z_j^2(t)}\,\cos\theta_j(t)\nonumber\\
&+&K_{c,12})N_j\frac{z_i(t)\cos\theta_i(t)}{\sqrt{1-z_i^2(t)}}
\;.\nonumber\\ \end{eqnarray}

\section{Gross-Pitaevskii equations predictions: comparison with
ordinary differential equations results}

So far we have discussed how AJJs dynamics can be described by means of the ODEs,
i.e. Eqs. (\ref{singlecomplete}) and (\ref{twocomplete}). We know that AJJs
dynamics can be analyzed, in the mean-field approximation, in terms of partial
differential equations, i.e. the GPEs. This description can be achieved
proceeding from the Heisenberg motion equations for the field operators $\hat{\Psi}_i({\bf
r},t)$, ($i=1,2$), associated to the Hamiltonian (\ref{sqhamiltonian}), that is
\beq
\label{eomm}
i \hbar \partial_{t} \hat{\Psi}_i=[\hat{\Psi}_i, \hat{H}] \;.
\eeq
The average - denoted by $\langle ...\rangle$ - of both sides of Eq.
(\ref{eomm})  evaluated with respect to the coherent state, provides the two coupled GPEs
\beq \label{GPE} i \hbar \frac{\partial \Psi_i}{\partial t} =
-\frac{\hbar^2}{2m_i}\nabla^2 \Psi_i+[V_{trap}({\bf r})
+g_{i}|\Psi_i|^2+g_{ij}|\Psi_j|^2]\Psi_i\; . \eeq
The macroscopic wave functions $\Psi_i({\bf r},t)=\langle
\hat{\Psi}_i({\bf r},t)\rangle$ of interacting BECs in the trapping potential
$V_{trap}({\bf r})$ at zero-temperature satisfy Eq. (\ref{GPE}). The wave function $\Psi_i({\bf r},t)$ is subject to the normalization condition
\beq
\label{normalizationwf}
\int  d^{3} {\bf r}\,
|\Psi_i({\bf
r},t)|^2=N_i \; .\eeq

\begin{figure}
\centering
\resizebox{\columnwidth}{!}{
\begin{tabular}{cc}
\epsfig{file=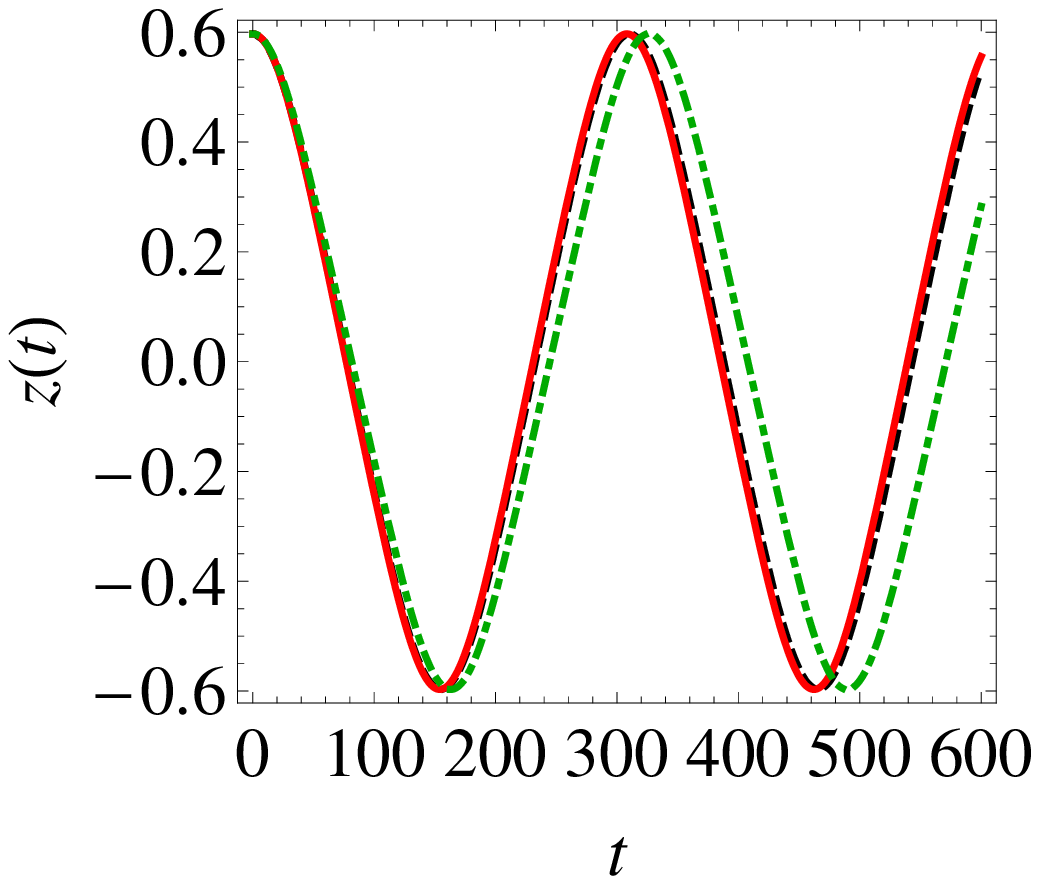,width=16cm,clip=}&
\epsfig{file=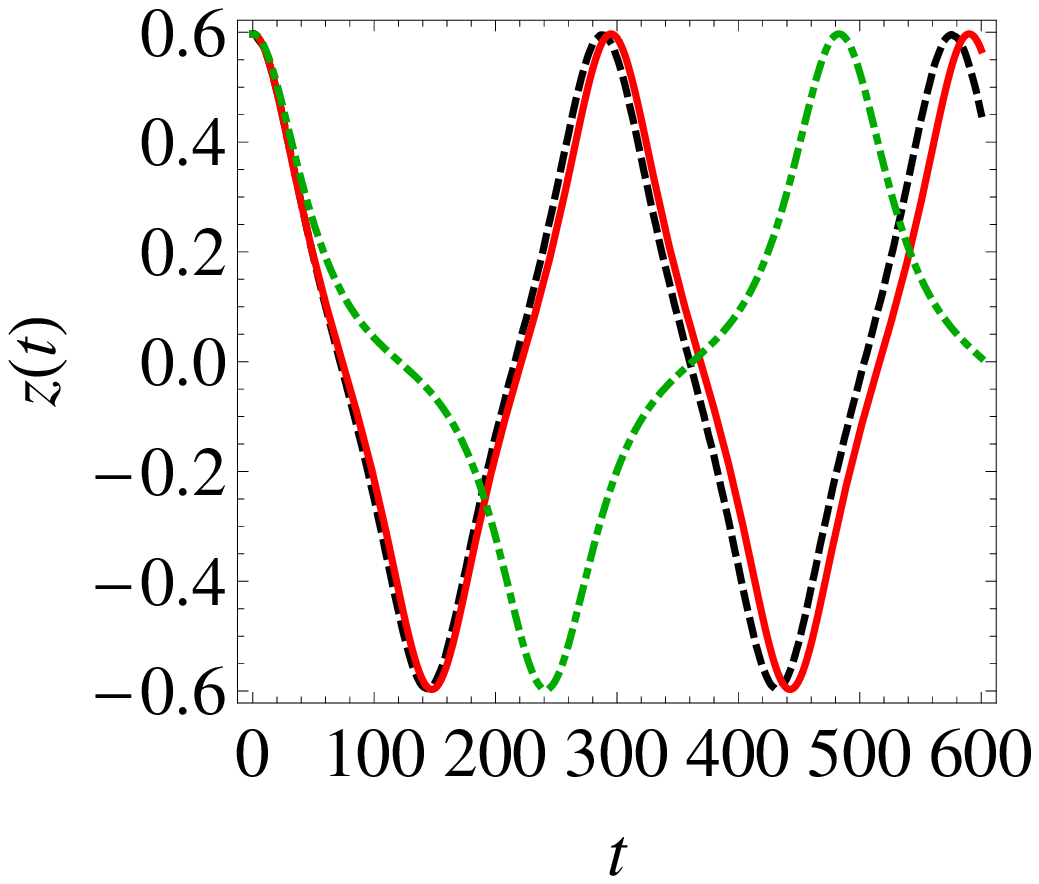,width=16cm,clip=}\\
\epsfig{file=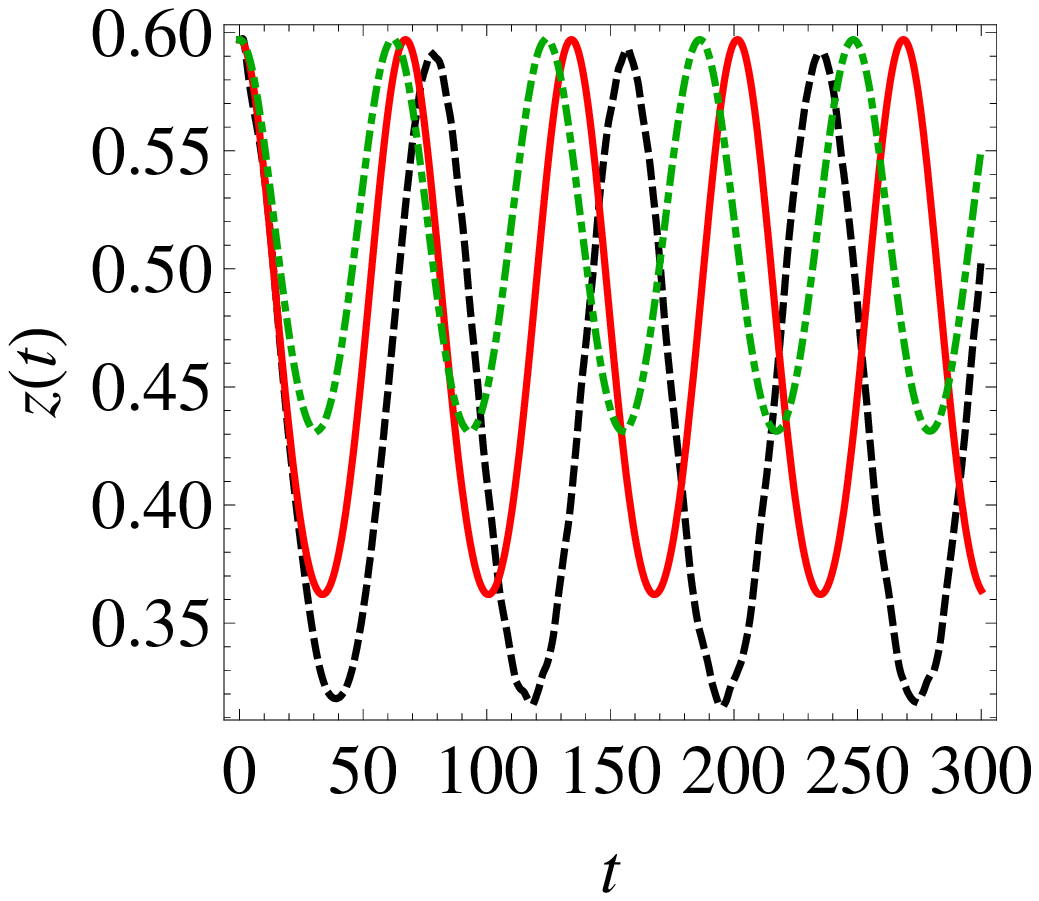,width=16cm,clip=}&
\epsfig{file=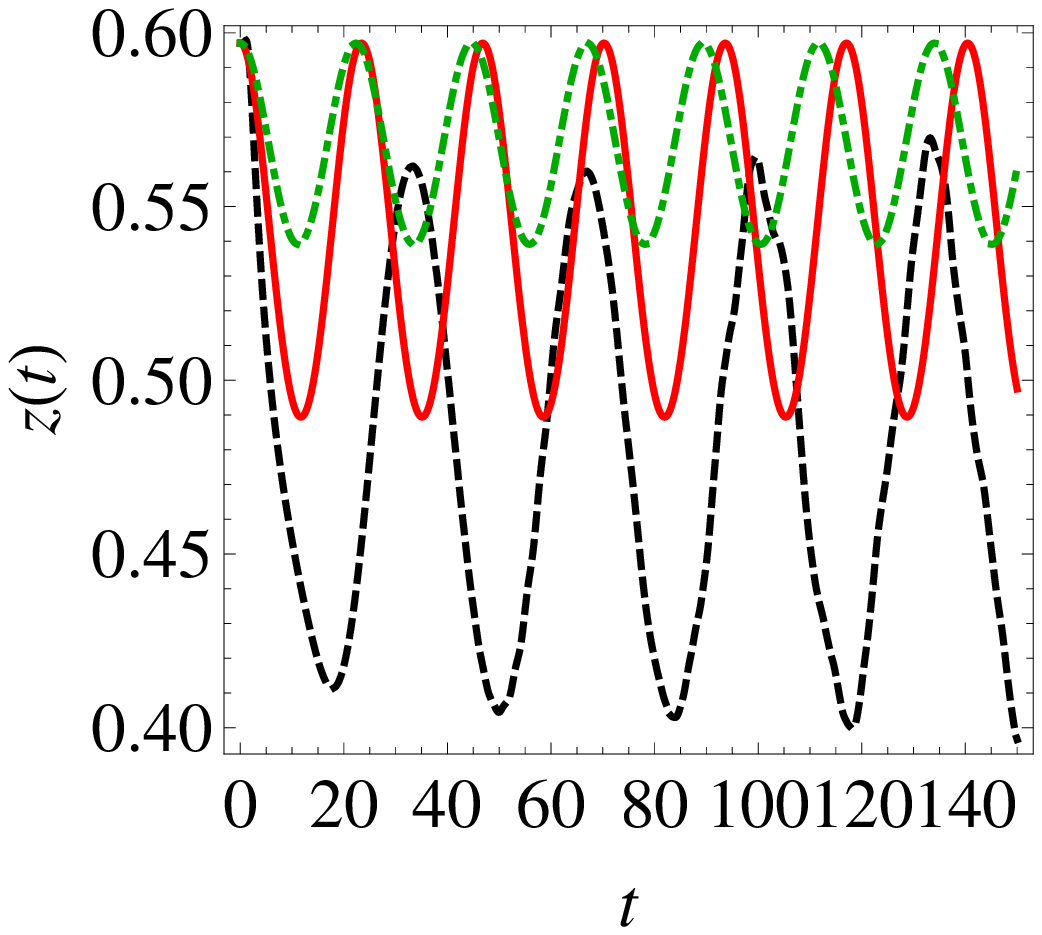,width=16cm,height=14cm,clip=}
\end{tabular}}
\caption{Fractional imbalance $z(t)$ vs. time for single component atomic
Josephson junctions. The parameters of the double-well potential (\ref{PT}) are chosen to be $b=1$
and $z_0=3$. The dashed line represents data from
the integration of GPE (\ref{single1dGPE}), the continuous line represents data from
the integration of ODEs (\ref{singlecomplete}), and the dot-dashed one represents data
from the  integration ODEs (\ref{singlecomplete})  with $K_{c}=V=0$. We have set
$N=200$ and $K=4.955\times10^{-3} $. We have used the initial conditions
$z(0)=0.6$ and $\theta(0)=0$. In the top panels (from left to right): $U=0.05\,K$,
$K_{c}=-1.842\times10^{-6}$, $V=2.268\times10^{-7}$; $U=0.1\,K$,
$K_{c}=-3.684\times10^{-6}$, $V=4.535\times10^{-7}$.
In the bottom panels (from left to right): $U=0.2\,K$,
$K_{c}=-7.368\times10^{-6}$, $V=9.070\times10^{-7}$; $U=0.5\,K$,
$K_{c}=-1.842\times10^{-5}$, $V=2.268\times10^{-6}$.
Time is measured in units of $\omega^{-1}$ and
energies are measured in units of $\hbar \omega$.}\label{fig2}
\end{figure}
\begin{figure}
\centering
\resizebox{\columnwidth}{!}{
\begin{tabular}{cc}
\epsfig{file=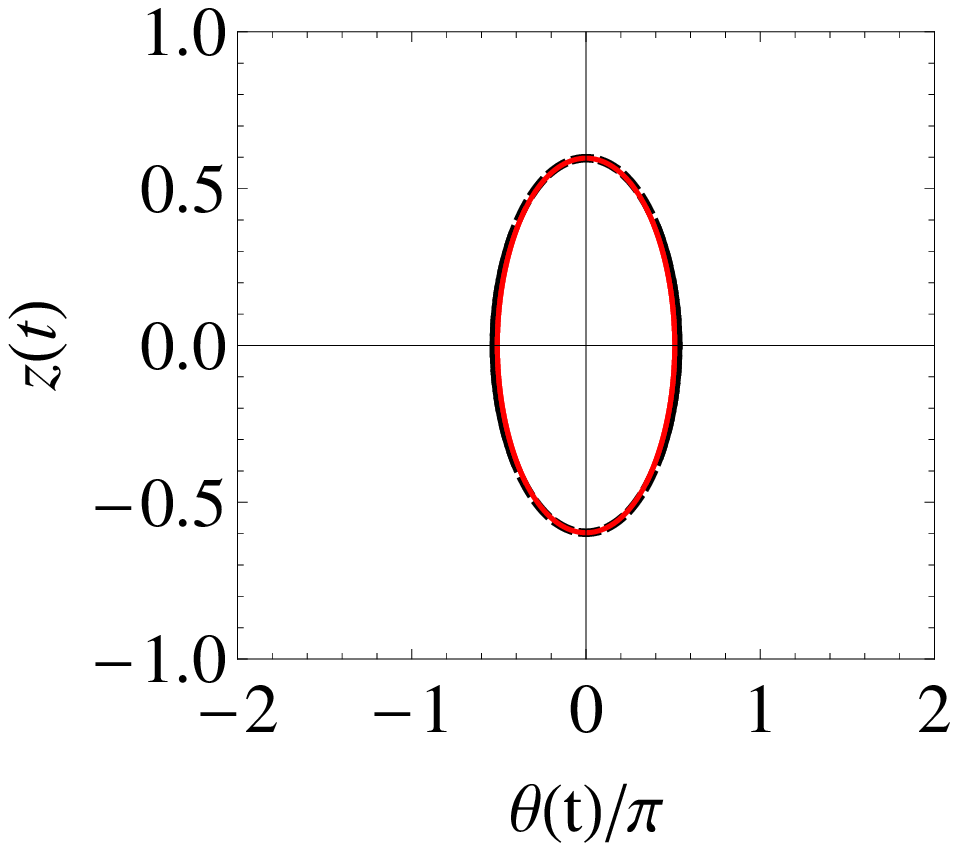,width=16cm,height=12cm,clip=}&
\epsfig{file=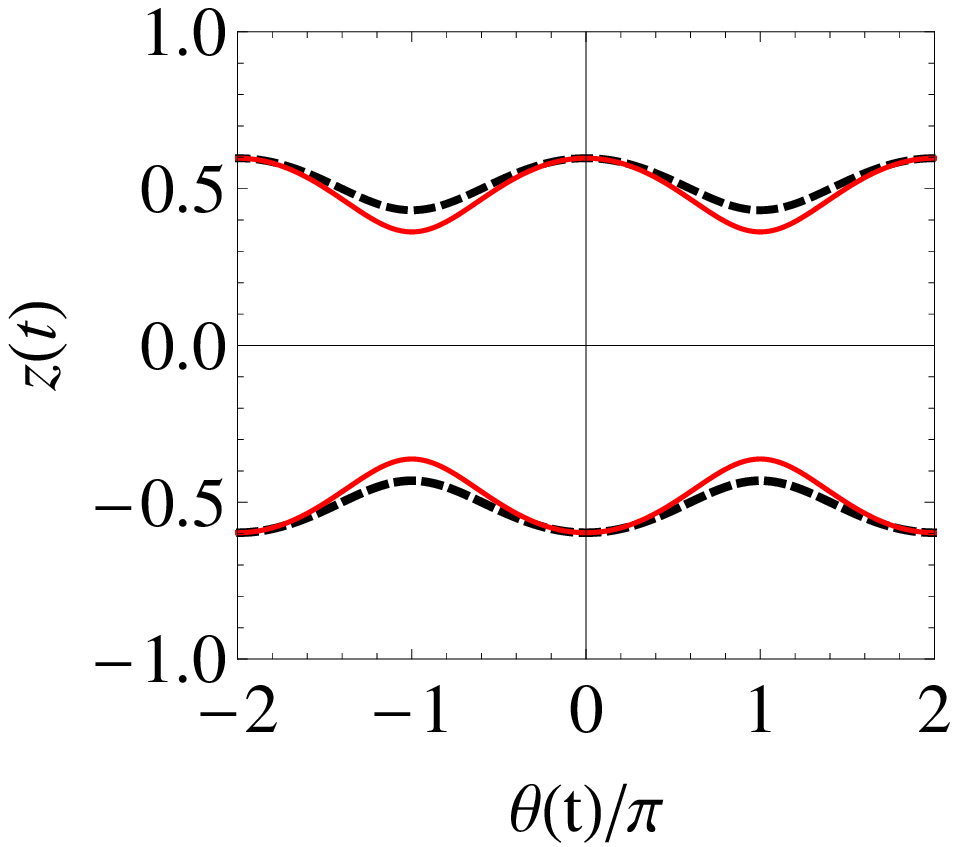,width=16cm,height=12cm,clip=}
\end{tabular}}
\caption{Phase diagrams of the fractional imbalance $z(t)$ vs. macroscopic phase
$\theta(t)$ for single component atomic Josephson junctions.
The parameters of the double-well potential (\ref{PT}) are the same as in Fig.
2.
In both the panels we have set $N=200$ and $K=4.955\times10^{-3}$. \\
Left panel: the dashed line represents data from the ODE
(\ref{singlecomplete}) with $U=0.05\,K$ and $K_c=V=0$; the continuous line
represents data from the ODE
(\ref{singlecomplete}) with $U=0.05\,K$, $K_c=-1.842
\times10^{-6}$ and $V=2.268 \times10^{-7}$.\\
The right panel shows the phase diagram for the self-trapping. In this panel the dashed line represents data from the ODE
(\ref{singlecomplete}) with $U=0.2\,K$ and $K_c=V=0$; the continuous line
represents data from the ODE
(\ref{singlecomplete}) with $U=0.2\,K$, $K_c=-7.368
\times10^{-6}$ and $V=9.070 \times10^{-7}$. Initial conditions are the
same as in Fig. 2. Time is measured in units of $\omega^{-1}$ and
energies are measured in units of $\hbar \omega$.}\label{fig3}
\end{figure}

\begin{figure}
\centering
\resizebox{\columnwidth}{!}{
\begin{tabular}{cc}
\epsfig{file=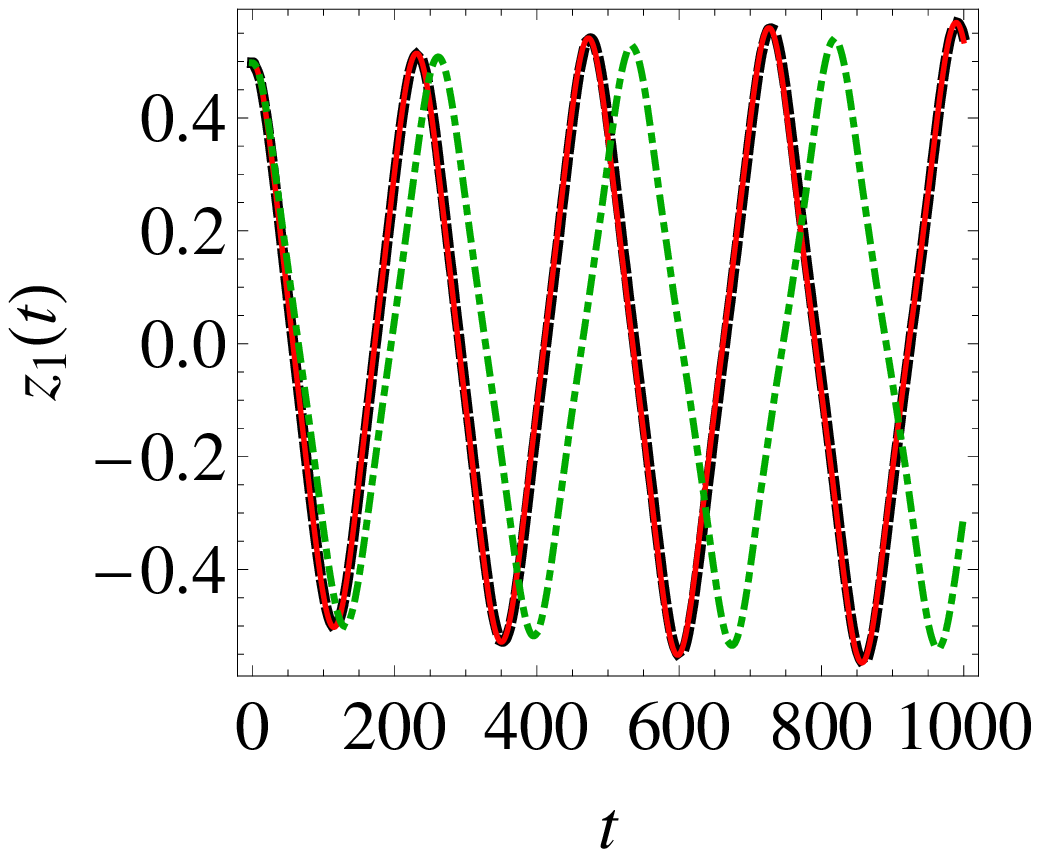,clip=}&
\epsfig{file=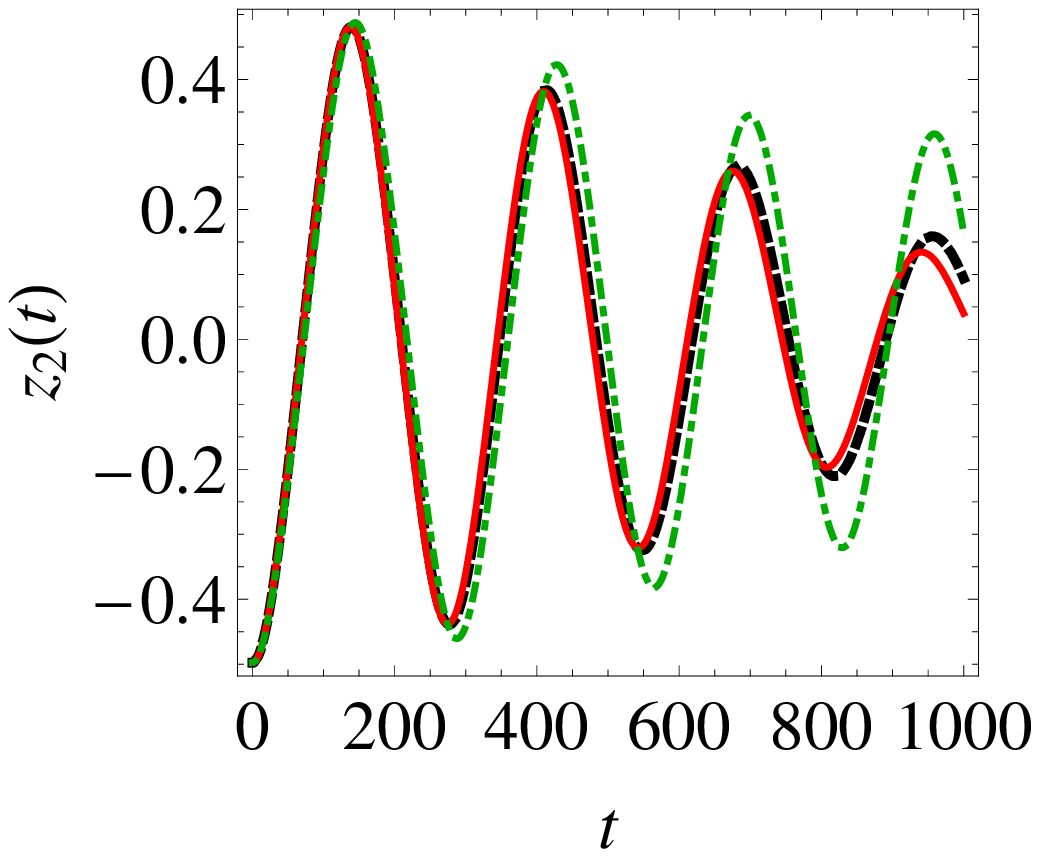,clip=}\\
\epsfig{file=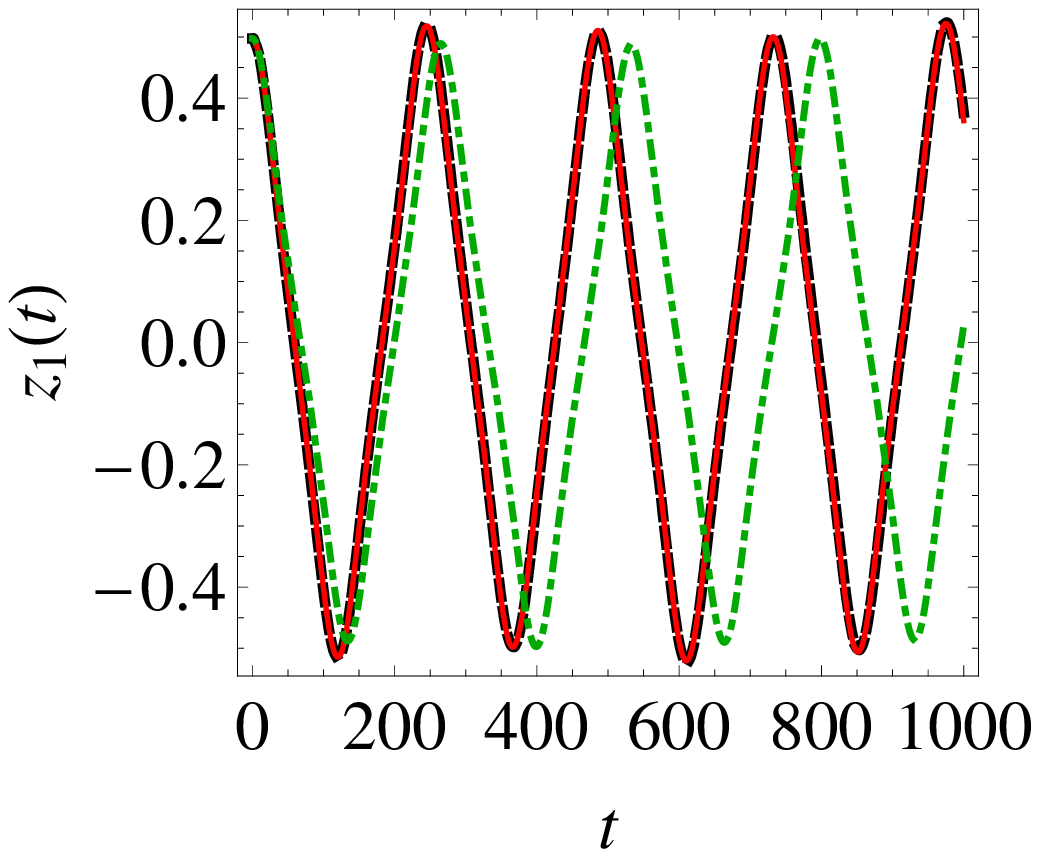,clip=}&
\epsfig{file=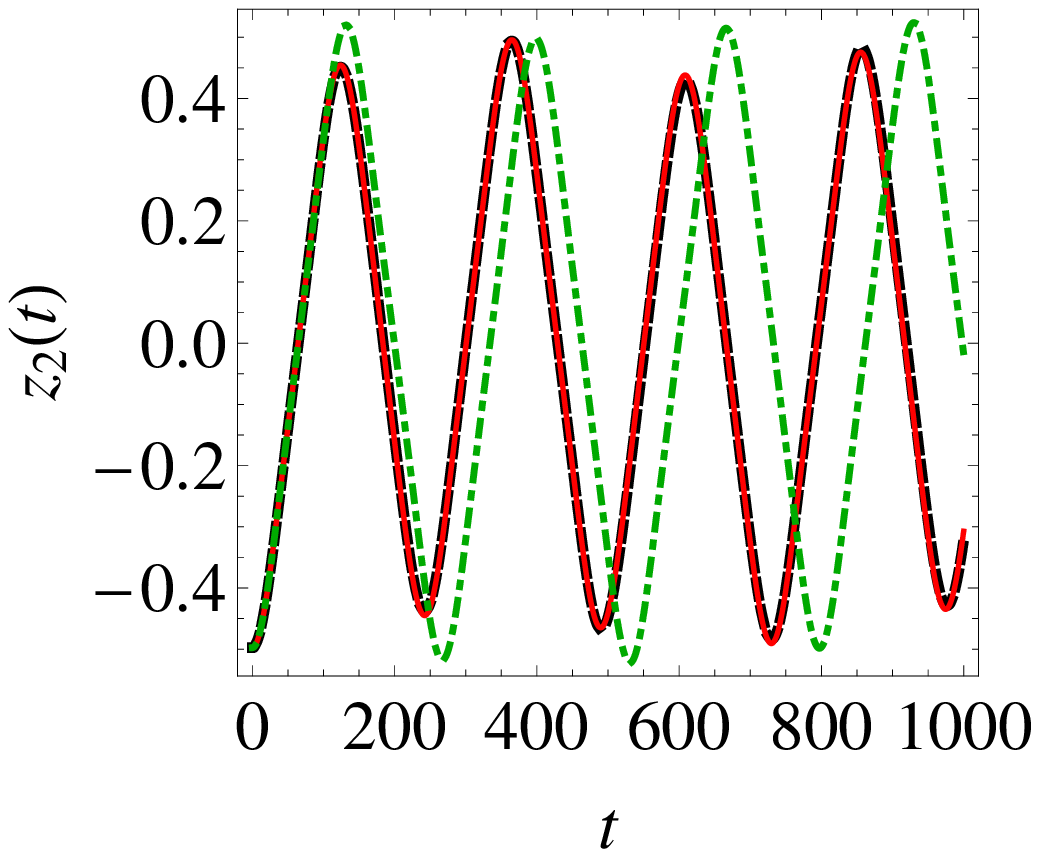,clip=}\\
\epsfig{file=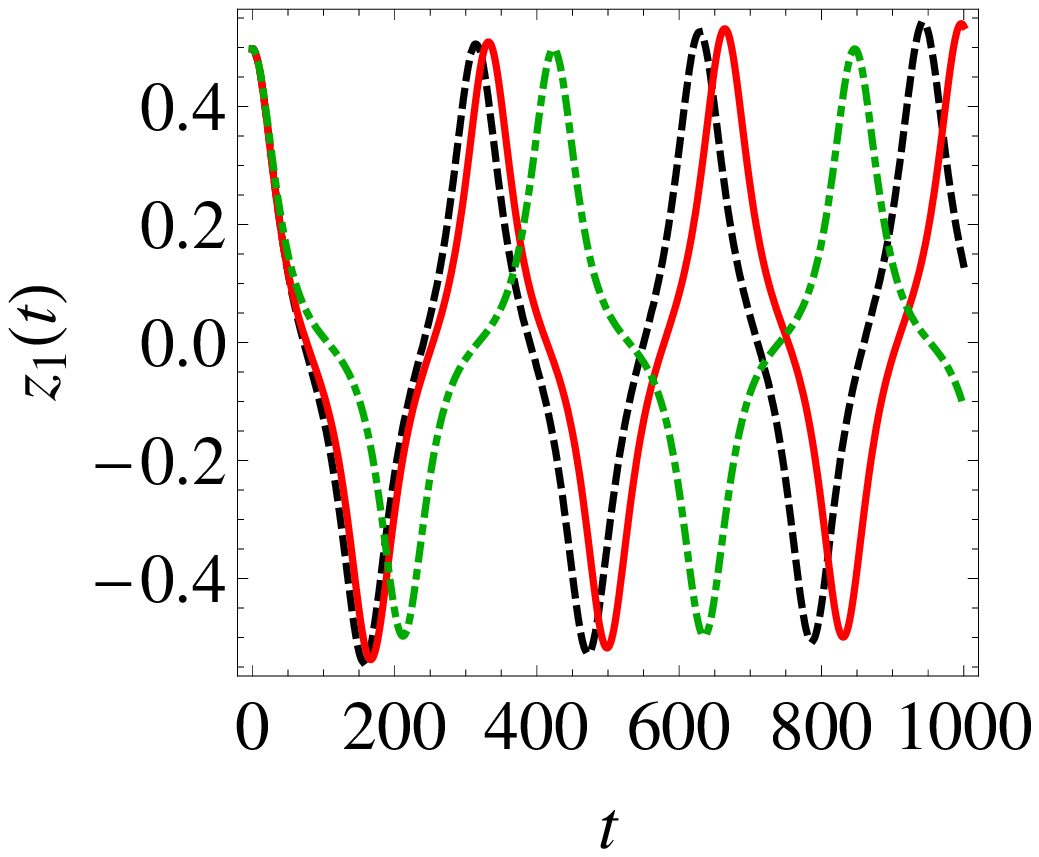,clip=}&
\epsfig{file=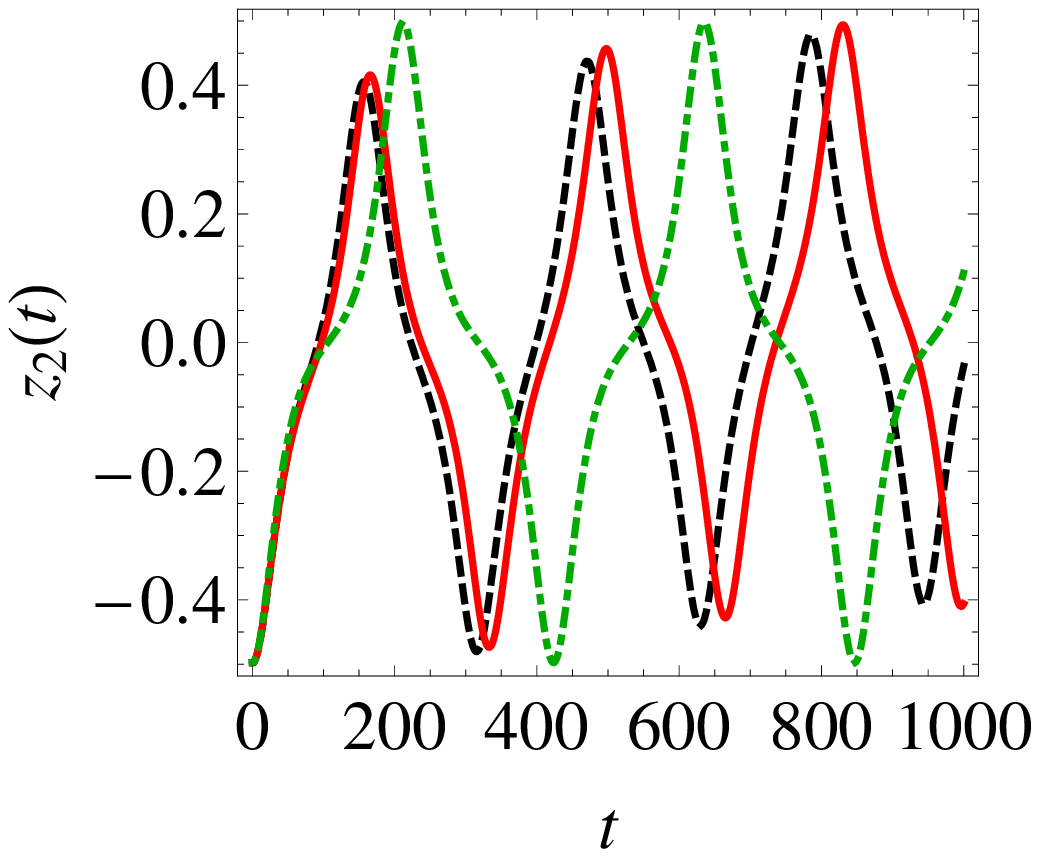,clip=}
\end{tabular}}
\caption{Fractional imbalance $z_i(t)$ of the two bosonic species vs. time.
The parameters of the double-well potential (\ref{PT}) are chosen to be $b=1$
and $z_0=3$. Here, the dashed line represents data from
the integration of GPEs (\ref{two1dGPE}), the continuous line represents data from
the integration of ODEs (\ref{twocomplete}), and the dot-dashed line represents data
from the  integration ODEs (\ref{twocomplete})  with $K_{c,i}=V_i=K_{c,12}=V_{12}=0$
(i.e. the ODEs analyzed in Ref. \cite{mazzarella}). We have fixed $N_1=200$ and $N_2=100$. Moreover,
$K_1=K_2 \equiv K=4.955\times10^{-3} $, $U_1=U_2\equiv U=0.1\,K$,
$K_{c,1}=K_{c,2}\equiv K_{c}=-3.684\times10^{-6}$,
$V_1=V_2 \equiv V=2.268\times10^{-7}$.
We used the initial conditions $z_1(0)=0.5=-z_2(0)$ and
$\theta_1(0)=\theta_2(0)=0$.  In the top panels we set
$U_{12}=-U/20$, $K_{c,12}=-K_{c}/20$, $V_{12}=-V/40$, in the middle panels
$U_{12}=-U/2$, $K_{c,12}=-K_{c}/2$, $V_{12}=-V/4$, and in the bottom panels
$U_{12}=-U$,  $K_{c,12}=-K_{c}$, $V_{12}=-V/2$. Time is measured
in units of $(\omega_1)^{-1}=(\omega_2)^{-1} \equiv \omega^{-1}$ and energies are measured in units of $\hbar \omega$.}\label{fig4}
\end{figure}

\begin{figure}
\centering
\resizebox{\columnwidth}{!}{
\begin{tabular}{cc}
\epsfig{file=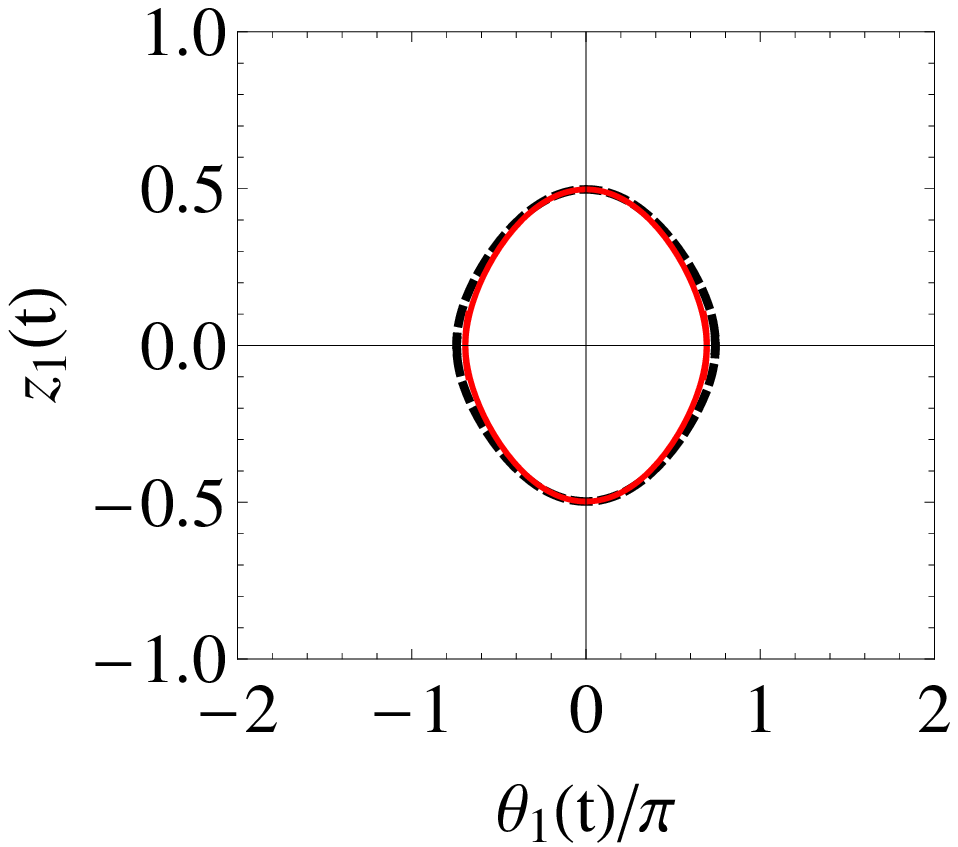,width=16cm,height=12cm,clip=}&
\epsfig{file=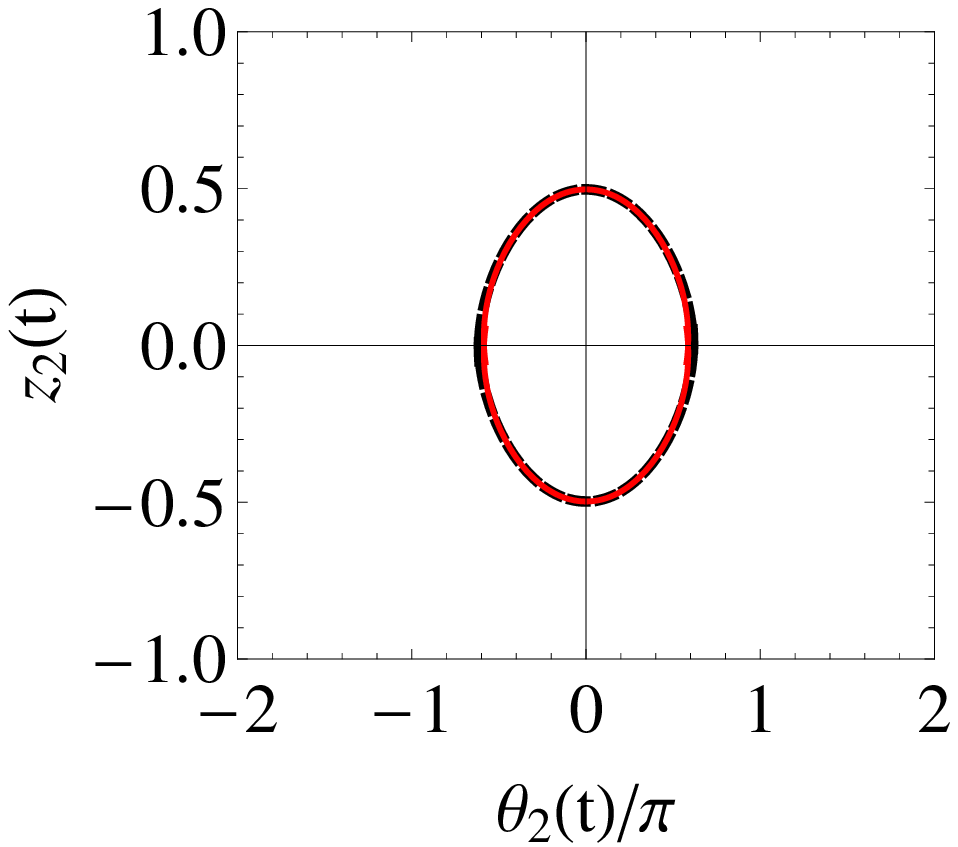,width=16cm,height=12cm,clip=}
\end{tabular}}
\caption{Phase diagrams of the fractional imbalance $z_i(t)$ vs. macroscopic
phase $\theta_i(t)$ of the two bosonic species.
The parameters of the double-well potential (\ref{PT}) are the same as in Fig.
4.
In both the panels we have set  $N_1=200$, $N_2=100$, $K_1=K_2 \equiv K=4.955\times10^{-3}$,
$U_1=U_2 \equiv U=0.1K$. In both the panels, the dashed line represents data from the ODEs
(\ref{twocomplete}) with $U_{12}=-U/2$ and  $K_{c,i}=V_i=K_{c,12}=V_{12}=0$, the continuous line
represents data from the ODEs with $U_{12}=-U/2$, $K_{c,1}=K_{c,2} \equiv
K_c=-3.684 \times10^{-6}$, $V_1=V_2 \equiv
V=2.268\times10^{-7}$, $K_{c,12}=-K_c/2$, $V_{12}=-V/4$. Initial conditions are
the same as in Fig. 4. Time is measured
in units of $(\omega_1)^{-1}=(\omega_2)^{-1} \equiv \omega^{-1}$ and energies are measured in units of $\hbar \omega$.}\label{fig5}
\end{figure}

\begin{figure}
\centering
\resizebox{\columnwidth}{!}{
\begin{tabular}{cc}
\epsfig{file=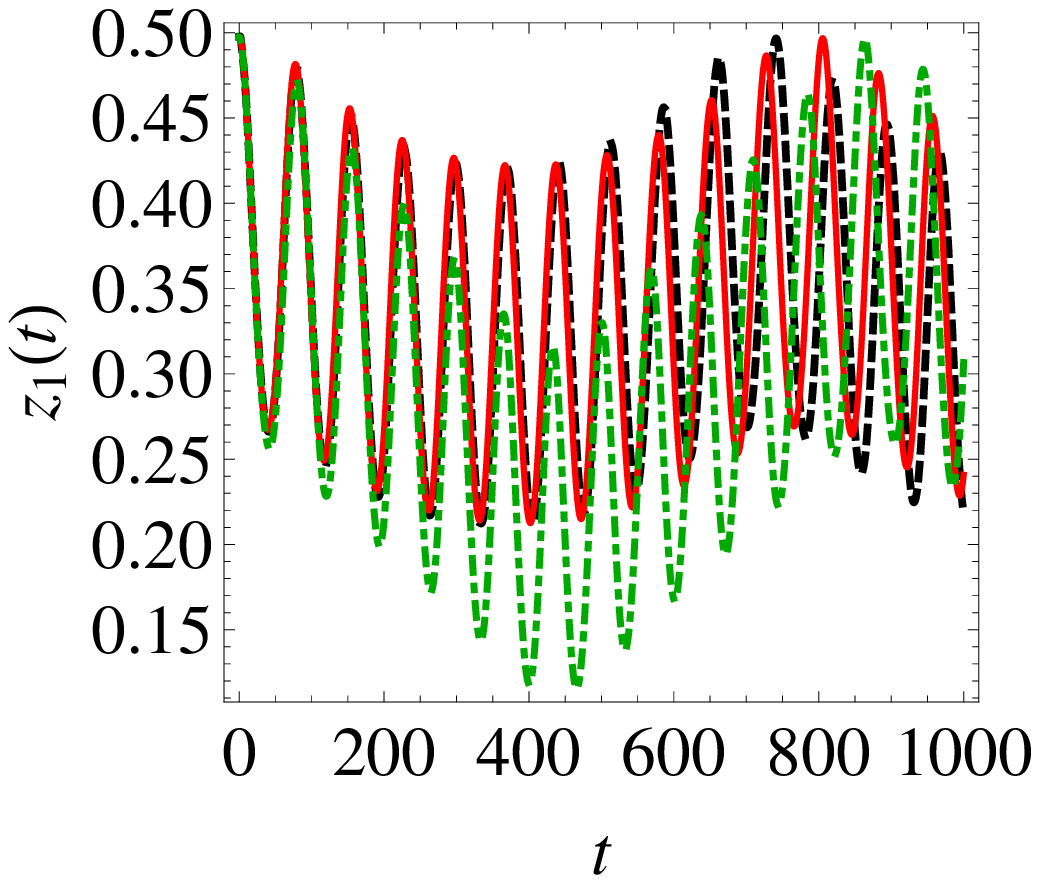,width=4cm,clip=}&
\epsfig{file=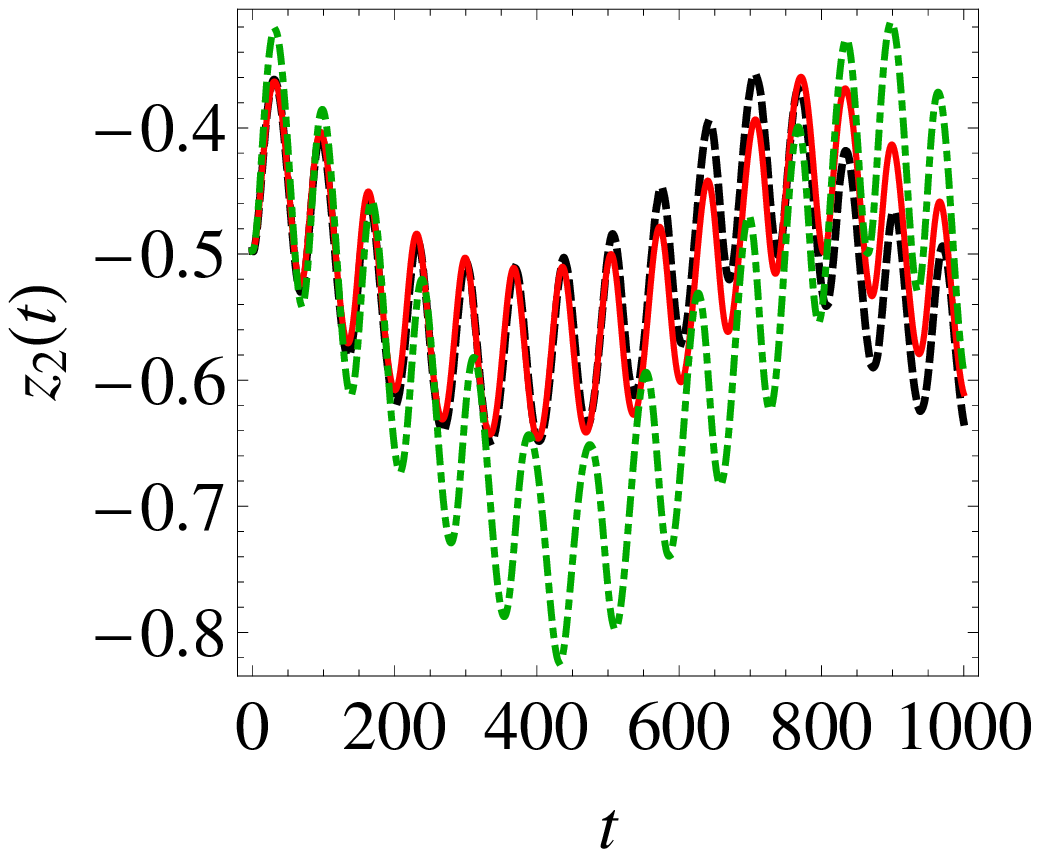,width=4cm,height=3.4cm,clip=}
\end{tabular}}
\caption{Fractional imbalance $z_i(t)$ of the two bosonic species vs. time. In
the double-well potential (\ref{PT}) we set $b=1$ and $z_0=3$.
In this figure the dashed line represents data from
the integration of GPEs (\ref{two1dGPE}), the continuous line represents data from
the integration of ODEs (\ref{twocomplete}), and the dot-dashed line represents data
from the  integration ODEs (\ref{twocomplete})  with  $K_{c,i}=V_i=K_{c,12}=V_{12}=0$ (i.e. the ODEs analyzed in Ref. \cite{mazzarella}). We have fixed
$N_1=200$ and $N_2=100$. Moreover, $K_1=K_2 \equiv K=4.955\times10^{-3} $,
$U_1=U_2\equiv U=0.1\,K$, $K_{c,1}=K_{c,2}\equiv K_{c}=-3.684\times10^{-6}$, $V_1=V_2 \equiv V=2.268\times10^{-7}$, $U_{12}=-2\,U$, $K_{c,12}=-2\,K_{c}$
$V_{12}=-V$. Initial conditions are the same as in Fig. 4. Time is measured
in units of $(\omega_1)^{-1}=(\omega_2)^{-1} \equiv \omega^{-1}$ and energies are measured in units of $\hbar \omega$.}\label{fig6}
\end{figure}

\begin{figure}
\centering
\resizebox{\columnwidth}{!}{
\begin{tabular}{cc}
\epsfig{file=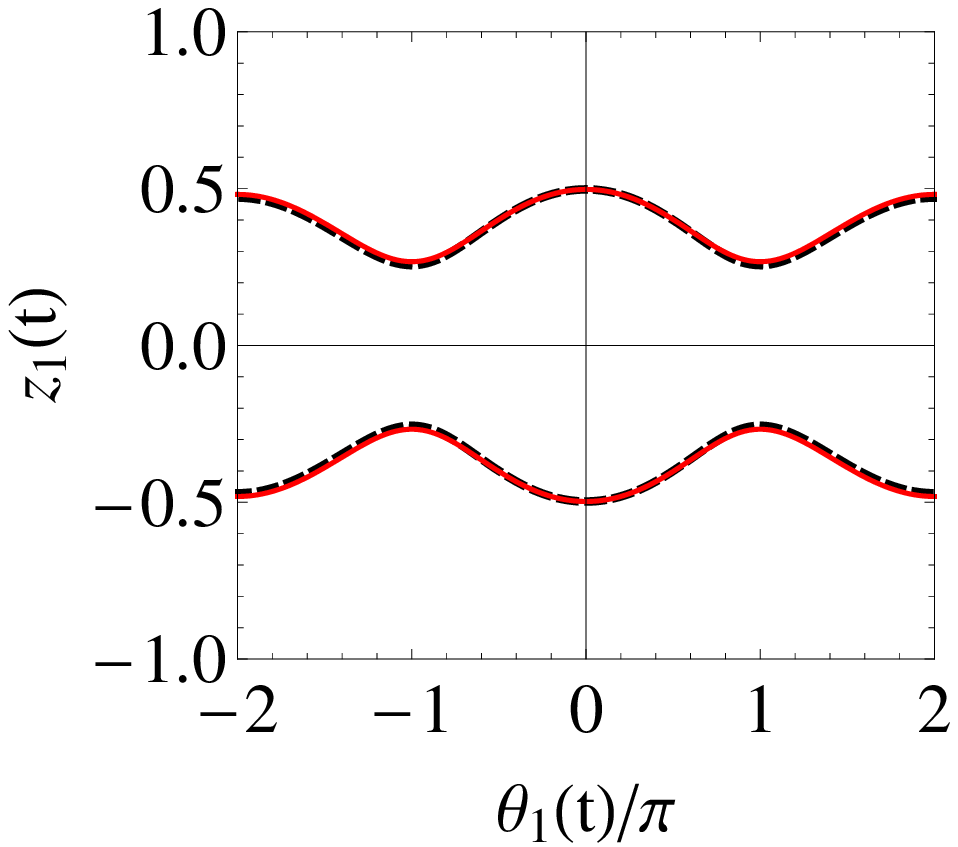,width=16cm,height=12cm,clip=}&
\epsfig{file=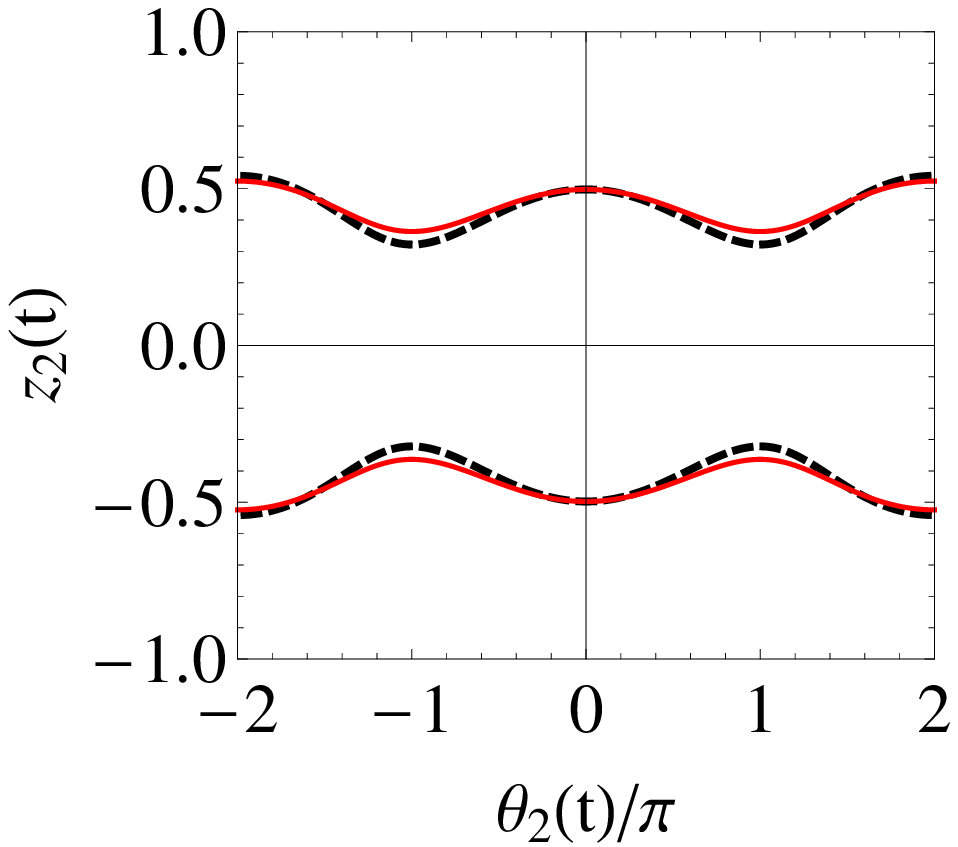,width=16cm,height=12cm,clip=}
\end{tabular}}
\caption{Phase diagrams of the fractional imbalance $z_i(t)$ vs. macroscopic
phase $\theta_i(t)$ of the two bosonic species for the self-trapping.
The parameters of the double-well potential (\ref{PT}) are the same as in Fig.
4.
In both the panels we have set  $N_1=200$, $N_2=100$, $K_1=K_2 \equiv K=4.955\times10^{-3}$,
$U_1=U_2 \equiv U=0.1\,K$. In both the panels, the dashed line represents data from the ODEs
(\ref{twocomplete}) with $U_{12}=-2\,U$ and  $K_{c,i}=V_i=K_{c,12}=V_{12}=0$, the continuous line
represents data from the ODEs with $U_{12}=-2\,U$, $K_{c,1}=K_{c,2} \equiv
K_c=-3.684 \times10^{-6}$, $V_1=V_2 \equiv
V=2.268\times10^{-7}$, $K_{c,12}=-2K_c$, $V_{12}=-V$. Initial conditions are the same
as in Fig. 4. Time is measured
in units of $(\omega_1)^{-1}=(\omega_2)^{-1} \equiv \omega^{-1}$ and energies are measured in units of $\hbar \omega$.}\label{fig7}
\end{figure}

We are interested to study the dynamical oscillations of the
populations of each
condensate between the left  and right  wells when the the barrier
is large enough so that the link is weak.
To exploit the strong harmonic confinement  in the ($x$-$y$)
plane and  get the effective one-dimensional (1D) equations describing the dynamics in
the $z$ directions,
we write the Lagrangian associated to the GPE equations
in (\ref{GPE})
\begin{eqnarray}
\label{lagrangian} L &=& \int d^3 {\bf
r}\,\bigg(\bigg[\sum_{i=1,2} \bar \Psi_i\big(i \hbar
\frac{\partial}{\partial t}+\frac{\hbar^2}{2m_i} \nabla^2
\big)\Psi_i\nonumber\\ &-&V_{trap}({\bf
r})|\Psi_i|^2-\frac{g_{i}}{2}|\Psi_i|^4\bigg]-g_{ij}|\Psi_i|^2|
\Psi_j|^2\bigg)\;,\nonumber\\
 \end{eqnarray}
where $\bar \Psi_i$ denotes the complex conjugate
of $\Psi_i$, and $i \neq j$; then, by following the decomposition
(\ref{decomposition}) and the Gaussian approximation
for the radial part of wave function, we adopt the ansatz
\beq
\label{ansatzb} \Psi_i(x,y,z,t)=\frac{1}{\sqrt{\pi} a_{\bot,i}}
\exp \bigg[-\frac{x^2+y^2}{2 a_{\bot, i}^2}\bigg] f_{i}(z,t)\;
,\eeq
where the field $f_i(z,t)$ obey to $\int_{-\infty}^{+\infty}dz|f_i(z)|^2=N_i$, so that the normalization condition given
by Eq. (\ref{normalizationwf}) is satisfied. Note that the Gaussian ansatz with the transverse width simply given by $a_{\bot,i}$ is reliable under
very strong transverse confinements, namely when $\tilde g_i |f_i|^2 \ll 2\,\hbar \omega_{i}$ \cite{salasnichvecsol}. By inserting the ansatz (\ref{ansatzb})
in Eq. (\ref{lagrangian}) and performing the integration in the radial
plane, we obtain the effective 1D Lagrangian for the field $f_i(z,t)$. Such an effective 1D Lagrangian reads
\begin{eqnarray}
\label{effectivelagrangian} \tilde L &=& \int d
z\,\bigg(\bigg[\sum_{i=1,2}
\bar f_i\big(i \hbar \frac{\partial}{\partial
t}+\frac{\hbar^2}{2m_i}\frac{\partial^2}{\partial z^2} \big)f_i \nonumber\\
&-& (\epsilon_i+V_{DW}(z))|f_i|^2-
\frac{\tilde g_{i}}{2}|f_i|^4\bigg]- \tilde g_{ij}|f_i|^2|f_j|^2\bigg)\;,\nonumber\\
\end{eqnarray}
where $\epsilon_i$ is given by
$\displaystyle{\epsilon_i=\frac{\hbar^2}{2m_i
a_{\bot,i}^2}+\frac{m_i \omega_i^2 a_{\bot,i}^2}{2}} $.
By varying $\tilde L$ with respect
to $\bar f_i$, we obtain the 1D GPE for the field $f_i$
\begin{equation}
\label{two1dGPE}  i \hbar \frac{\partial f_i}{\partial t} =
-\frac{\hbar^2}{2m_i}\frac{\partial^2 f_i}{\partial
z^2}+[\epsilon_i+V_{DW}(z) + \tilde g_{i}|f_i|^2+ \tilde
g_{ij}|f_j|^2]f_i \; .\end{equation}
In the presence of a single bosonic component, $g_{12}=0$; then, the two coupled 1D GPEs Eq. (\ref{two1dGPE}),
omitting the species index $i$, reduce to
\begin{equation}
\label{single1dGPE}  i \hbar \frac{\partial f}{\partial t} =
-\frac{\hbar^2}{2m}\frac{\partial^2 f}{\partial
z^2}+[\epsilon+V_{DW}(z) + \tilde g|f|^2]f \; .\end{equation}
Now, we observe that it is possible to write the fields $f_i$, ($i=1,2$), by using the two-mode
approximation as done, for example, in Ref. \cite{mazzarella}
\begin{eqnarray}
&&f_i(z,t)=\psi_{L,i}(t)\phi_{L,i}(z)+\psi_{R,i}(t)\phi_{R,i}(z)\nonumber\\
&& \psi_{\alpha,i}(t)=\sqrt{N_{\alpha,i}(t)}\exp(i\theta_{\alpha,i}(t))
\;,\end{eqnarray}
with $\phi_{\alpha,i}(z)$ constructed as discussed in Sec. III, see Eq. (\ref{finaldecomposition}).
Then, one takes into account the overlaps both between $\phi_{\alpha}$'s localized in the same well and between $\phi_{\alpha}$'s localized in
different wells. By following the same path as in Ref. \cite{mazzarella}, when
the inter-species coupling constant $g_{12}$ is finite, it is possible to recover  Eqs. (\ref{twocomplete}) for binary AJJs, while for
$g_{12}$ equal to zero one gets back the Eqs. (\ref{singlecomplete}) for single component AJJs .

At this point - both for single component
and for two components AJJs - we may compare  the predictions of
the ODEs, Eqs. (\ref{singlecomplete}) and (\ref{twocomplete}), and those of
the GPEs, Eqs. (\ref{single1dGPE}) and (\ref{two1dGPE}).
The results of this analysis are reported in Fig. 2 for the
single component case, and in Figs. 4, 6 for the two components case.
In obtaining Fig. 2 we have fixed the parameters $b$ and $z_0$ of the
double-well potential (\ref{PT}). Then, by using the functions
(\ref{finaldecomposition}) into the third of Eqs. (\ref{selfparameters}), we have
obtained the tunneling amplitude $K$. We have keeped fixed $K$ and we have
plotted the predictions of the ODEs (\ref{singlecomplete}) for $z(t)$ in
correspondence to different intra-species interactions both when $K_c$ and $V$ are zero - dot-dashed lines -  and in the presence of $K_c$
and $V$ - continuous lines; the dashed lines represent $z(t)$ obtained by
numerically integrating the GPE (\ref{single1dGPE}). In Figs. 4, 6 we have fixed the tunneling amplitude
$K_i$ - as done previously in the single component case - and the intra-species
interaction $U_i$, and we have plotted the
predictions of the ODEs (\ref{twocomplete}) for $z_i(t)$ in correspondence to different
inter-species interactions both when $K_{c,i}$, $V_i$, $K_{c,12}$, $V_{12}$ are
all equal to zero - dot-dashed lines - and in the
presence of $K_{c,i}$, $V_i$, $K_{c,12}$, $V_{12}$ - continuous lines; again, the dashed
lines represent $z_i(t)$ obtained by numerically integrating the GPEs (\ref{two1dGPE}). In the two top panels of Fig. 2
and in all the panels of Fig. 4, we have plotted the temporal evolution of the
bosonic fractional imbalances $z$ when they oscillate around a zero time-averaged
value, i.e. $\langle z(t)\rangle = 0$. We see that the usually-neglected
nonlinear terms play a crucial role
in order to improve the agreement between the GPEs the ODEs
predictions. In fact, neglecting these terms, the solutions of ODEs and GPEs
diverge rather rapidly, as shown by dot-dashed lines in Fig. 2 - single
component AJJs - and by dot-dashed lines in Fig. 4, for two components AJJs.
The two bottom panels of Fig. 2 show the results of our
analysis when the intra-species interaction amplitude $U$ is
sufficiently large to induce oscillations of $z(t)$ around a non zero time-averaged
value, that is the self-trapping. We see that the inclusion within the description
of the system of the usually-neglected nonlinear terms produces an improvement in the agreement between the
ODEs and GPE predictions. In the two components case, from Fig. 4 we can see that the
nonlinearity associated to the intra-species interaction is not strong enough to induce oscillations of $z_i$ around a non zero time-averaged value. Nevertheless, if the inter-species interaction is sufficiently
large, oscillations of $z_i$ around $\langle z_i(t)\rangle
\neq 0$ are observed. We have reported this kind of behavior for both the
components in Fig. 6. From
this figure we can see that, especially in the case of large inter-species interaction, the role
played by the parameters describing the overlaps between $\phi_{\alpha}$'s localized in different wells becomes
essential to improve the agreement between the ODEs and the GPEs predictions.
Moreover, in Fig. 3 - single component AJJs -
and in Fig. 5 and Fig. 7 - two components AJJs - we show the phase-plain
portraits of the dynamical variables $z_i$ and $\theta_i$ for different values
of the macroscopic parameters (\ref{selfparameters}) and
(\ref{12selfparameters}) (see Figs. 3, 5, 7 for the details).
These figures show the comparison between the
trajectories in the phases space obtained by integrating the ODEs in the absence of the
usually-neglected nonlinear terms (dashed lines), and the trajectories obtained
from the improved version of ODEs (continuous lines). In particular, the left
and the right panels of Fig. 3 show the phases space trajectories for the Josephson and
self-trapping regimes, respectively, for single component AJJs. For two
components AJJs, in Fig. 5  we have plotted the phases space trajectories for the Josephson regime, and in
Fig. 7 we have plotted the phases space trajectories when the system is self-trapped.
From Figs. 3, 5, 7 we can see that the trajectories
predicted when the ODEs Eqs. (\ref{singlecomplete}) and (\ref{twocomplete}) are solved in the absence of the usually-neglected nonlinear terms are
sufficiently close to those predicted when these ODEs are solved in the
presence of the aforementioned terms. Then, the dynamical evolution predicted by the standard ODEs
reveals to have a good degree of reliability.

\section{Conclusions}
We have analyzed atomic Josephson junctions for a single
Bose gas and for binary mixtures of bosons in a
double-well  potential along the axial direction and a strong harmonic
confinement in the transverse directions.
We have shown that for both the cases the Hamiltonian
belongs to the extended Bose-Hubbard model and besides the density-density
interaction it contains the pair hopping and collisionally induced hopping terms.
These terms derive from the overlaps between wave functions localized in different
potential wells. We started from these Hamiltonian models and established
connections with spin Hamiltonians.
Proceeding from these, we have discussed the possibility to discriminate, under certain
conditions, different dynamical regimes sustained by the bosonic junctions. From the mean field analysis of the equations of motion for the
single-particle operators involved in the extended Bose-Hubbard
Hamiltonians, we have obtained the
ordinary differential equations that control the macroscopic dynamics of the atomic Josephson junctions.
Within the analysis of the atomic Josephson junctions macroscopic dynamics we
have plotted the phase-plane portraits of the dynamical variables (fractional
imbalance-relative phase) showing that the inclusion of  the aforementioned
collisionally induced hopping and pair hopping terms are crucial to get good agreement between the
dynamics of the Josephson model described by ordinary differential equations and the one of the time dependent Gross-Pitaevskii equations, especially when the atom-atom
interaction is strong.

Finally, it is important to remark that the obtained results are of
general validity also for more confining (e.g. not saturating to
zero at large distances) double-well potentials. Nevertheless, it is possible to
design a model of pair hopping and collisionally induced hopping for bosonic
atoms that is physically meaningful when optical lattices play the role of confining potentials.
Physical effects related to pair hopping and collisionally induced hopping
should be observable in generalizations of current experiments to detect
the superfluid and insulating phases \cite{eckholt}.\\

This work has been partially supported by Fondazione CARIPARO
through the Project 2006: "Guided solitons in matter waves and
optical waves with normal and anomalous dispersion". G. M. thanks A. B. Kuklov
and B. V. Svistunov for useful comments.




\section*{References}
\begin{thebibliography}{10}

\bibitem{einstein}
S. N. Bose, Z. Phys. {\bf 26}, 178 (1924); A.Einstein, Sitzungsber. K. Preuss.
Akad. Wiss., Phys. Math. K1. {\bf 22}, 261 (1924).

\bibitem{anderson}
M. H. Anderson, M. R. Matthews, C. E. Wieman, and. E. A. Cornell, Science {\bf 269},
198 (1995); K. B. Davis, M. O. Mewes, M. R. Andrews, N. J. van Druten, D. S. Durfee,
D. M. Kurn, and W. Ketterle, Phys. Rev. Lett. {\bf 75}, 3969 (1995);  C. C.
Bradley, C. A. Sackett, J. J. Tollett, and R. G. Hulet, {\it ibid.} {\bf 75}, 1687
(1995).

\bibitem{leggett} A. J. Leggett and F. Sols, Found. Phys.
{\bf 21}, 353 (1991); I. Zapata,  F. Sols, and A. J. Leggett, Phys. Rev. A   {\bf 57}, R28 (1998).

\bibitem{leggett01} A. J. Leggett, Rev. Mod. Phys.  {\bf 73}, 307 (2001).

\bibitem{smerzi} A. Smerzi, S. Fantoni, S. Giovanazzi, and S. R. Shenoy,
Phys. Rev. Lett.  {\bf 79}, 4950 (1997); S. Raghavan, A. Smerzi, S.Fantoni, and
S. R. Shenoy, Phys. Rev. A {\bf 59}, 620 (1999).

\bibitem{salasnich} L. Salasnich, A. Parola, L. Reatto,
J. Phys. B: At. Mol. Opt. Phys. {\bf 35}, 3205-3216 (2002).

\bibitem{salerno} M. Salerno,
Laser Phys. {\bf 4},  620-625 (2005).

\bibitem{book-barone} A. Barone and G. Patern\`{o},
{\it Physics and Applications of the Josephson effect} (Wiley, New
York, 1982); H. Otha, in {\it SQUID: Superconducting Quantum
Devices and their Applications}, edited by H.D. Hahlbohm and H.
Lubbig (de Gruyter, Berlin, 1977).

\bibitem{albiez} M. Albiez, R. Gati, J. F\"{o}lling, S.
Hunsmann, M. Cristiani, M. K. Oberthaler, Phys. Rev. Lett.  {\bf
95}, 010402 (2005).

\bibitem{gati} R. Gati and M. K. Oberthaler, J. Phys. B: At. Mol. Opt.
Phys.  {\bf 40}, R61-R89 (2007).

\bibitem{milburn} C. J. Milburn, J. Corney, E. M. Wright, and
D. F. Walls, Phys. Rev. A  {\bf 55}, 4318  (1997).

\bibitem{minardi} G. Thalhammer, G. Barontini, L. De Sarlo, J. Catani, F.
Minardi, and M. Inguscio, Phys. Rev. Lett.  {\bf 100}, 210402 (2008).

\bibitem{papp} S. B. Papp and C. E. Wieman,  Phys. Rev. Lett.  {\bf 97}, 180404 (2006).







\bibitem{xu} X. Xu, L. Lu, Y. Li, Phys. Rev. A  {\bf 78}, 043609 (2008).

\bibitem{satja} I. I. Satija, P. Naudus, R. Balakrishnan, J. Heward, M. Edwards,
C.W. Clark, Phys. Rev. A  {\bf 79}, 033616 (2009).

\bibitem{diaz1} B. Julia-Diaz, M. Guilleumas, M. Lewenstein, A. Polls, A.
Sanpera, Phys. Rev. A {\bf 78}, 023616 (2009).

\bibitem{mazzarella} G. Mazzarella, M. Moratti, L. Salasnich, M. Salerno and F. Toigo, J. Phys. B: At. Mol. Opt. Phys. {\bf 42}, 125301 (2009).

\bibitem{diaz2} B. Julia-Diaz, M. Mele-Messeguer,
M. Guilleumas, and A. Polls, Phys. Rev. A {\bf 80}, 043622 (2009).

\bibitem{wang} C. Wang, P. G. Kevrekidis, N. Whitaker and B. A. Malomed,
Physica D {\bf327}, 2922-2932 (2008).



\bibitem{mazz} G.Mazzarella, S. M. Giampaolo, F. Illuminati, Phys. Rev. A
{\bf 76}, 013625 (2006).

\bibitem{amico} L. Amico, G.Mazzarella, S. Pasini, F. S. Cataliotti, New J.
Phys, {\bf 12}, 013002 (2010).

\bibitem{svistunov} A. B. Kuklov and B. V. Svistunov, Phys. Rev. Lett.
{\bf 90}, 100401 (2003).

\bibitem{giampaolo} P. Buonsante, S. M. Giampaolo, F. Illuminati, V. Penna, A.
Vezzani, Phys. Rev. Lett. {\bf 100}, 240402 (2008); P. Buonsante, S. M. Giampaolo, F. Illuminati, V. Penna, A.
Vezzani, Eur. Phys. J. B {\bf 68}, 427 (2009).

\bibitem{roscilde} T. Keilmann, J. I. Cirac, T. Roscilde, Phys. Rev. Lett. {\bf
102}, 255304 (2009).

\bibitem{ferrini} G. Ferrini, A. Minguzzi, F. W. Hekking,   Phys. Rev. A
{\bf 78}, 023606(R) (2008).

\bibitem{averin} D. V. Averin, T. Bergeman, P. R. Hosur, and C. Bruder, Phys. Rev. A
{\bf 78}, 031601(R) (2008).

\bibitem{ananikian} D. Ananikian and T. Bergeman, Phys. Rev. A, {\bf 73}, 013604
(2006).

\bibitem{landau} L. Landau and L. Lifshitz,
{\it Course in Theoretical Physics}, Vol. 3, {\it Quantum
Mechanics: Non-Relativistic Theory}, (Pergamon, New York, 1959).


\bibitem{definition} This definition of $\hat{J}_{x}$ and $\hat{J}_{z}$
follows that of Refs. \cite{leggett01,ananikian}. A definition with exchanged $\hat{J}_{x}$ and $\hat{J}_{z}$ is also
widely used in literature - see e.g. \cite{ferrini} - and in theoretical quantum
optics textbooks - see e.g. \cite{barnett}.

\bibitem{barnett} S. M. Barnett and P. M. Radmore,
{\it Methods in Theoretical Quantum Optics}, (Oxford University Press, New York,
1997).



\bibitem{svistunovp} A. B. Kuklov and B. V. Svistunov, e-mail communications.

\bibitem{salasnichvecsol}
L. Salasnich, A. Parola, and L. Reatto, Phys. Rev. A {\bf 65}, 043614 (2002);
L. Salasnich, A. Parola, and L. Reatto, Phys. Rev. A {\bf 70}, 013606 (2004);
L. Salasnich and B. A. Malomed, Phys. Rev. A {\bf 74}, 053610
(2006).

\bibitem{eckholt} M. Eckholt and J. J. Garc\'{i}a Ripoll, Phys. Rev. A {\bf 77},
063603 (2008).









\end{thebibliography}
\end{document}